\documentstyle[preprint,aps,epsf]{revtex} 

\newcommand{\bseq}{\begin{equation}}
\newcommand{\eseq}{\end{equation}}

\newcommand{\beq}{\begin{equation} \bigskip}
\newcommand{\eeq}{\bigskip \end{equation}}

\newcommand{\beqa}{\begin{eqnarray} \bigskip}
\newcommand{\eeqa}{\bigskip \end{eqnarray}}

\mathchardef\curlyC="3243
\mathchardef\curlyD="3244
\mathchardef\curlyF="3246
\mathchardef\curlyG="3247
\mathchardef\curlyH="3248
\mathchardef\curlyI="3249
\mathchardef\curlyS="3253
\mathchardef\curlyU="3255
\mathchardef\curlyW="3257
\mathchardef\wiggle="3218

\def\CM{\curlyC}
\def\DM{\curlyD}
\def\FM{\curlyF}
\def\GM{\curlyG}
\def\HM{\curlyH}
\def\IM{\curlyI}
\def\SM{\curlyS}
\def\UM{\curlyU}
\def\WM{\curlyW \!}

\def\lra{\leftrightarrow}
\def\sss{\scriptscriptstyle}

\def\A{{\sss \! A}}
\def\a{{\sss \! a}}
\def\asym{{\sss \rm asymptotic}}
\def\asym{{r > r_{\rm c}}}
\def\AB{{\sss \! AB}}
\def\ABAB{{\sss \! AB \lra AB}}
\def\ABCD{{\sss \! AB \lra CD}}
\def\ABin{{\sss AB;{\rm in}}}
\def\ABQ{{\sss \! AB \lra Q}}
\def\ABR{{\sss \! AB \lra R}}
\def\ABtoR{{\sss \! AB \to R}}

\def\B{{\sss \! B}}
\def\BW{{\sss \! BW}}
\def\b{{\sss \! b}}
\def\C{{\sss \! C}}
\def\CD{{\sss \! CD}}
\def\CDin{{\sss CD;{\rm in}}}
\def\CDQ{{\sss \! CD \lra Q}}
\def\CDR{{\sss \! CD \lra R}}
\def\D{{\sss \! D}}
\def\ie{{\it i.e. }}
\def\FF{{\, {\bf F}_q \cdot {\bf F}_{\bar q} }}
\def\mR{{\rm M}_\R}
\def\naive{{na\" \i ve }}
\def\P{{\sss \! P}}
\def\q{{\sss \! q}}
\def\qbar{{\sss \! \bar q}}
\def\qq{{q\bar q}}

\def\qqqq{{qq\bar q\bar q}}
\def\Q{{\sss \! Q}}
\def\Qqq{{\sss \! Q_\qq}}
\def\QR{{\sss \! Q \lra R}}
\def\R{{\sss \! R}}
\def\RAB{{\sss \! R \to AB}}
\def\Rqq{{\sss \! R_\qq}}
\def\rAB{{r_{\AB}}}
\def\rqq{{r_{\qq}}}
\def\schrod{{Schr\" odinger }}
\def\sss{\scriptscriptstyle}
\def\sqroot#1{{\sqrt{#1}}}
\def\u{{\bf u}}
\def\z{{\bf z}}

\begin{document}


\title
{
\hfill \parbox[t]{2.5 in}
{\rm \small UMHEP 96--021}
\vskip 1 true cm
\bf The Multichannel Quark Model \\
}
\author
{
\bf John Weinstein \\
\it Department of Physics and Astronomy, \\
\it University of Mississippi, University, MS 38677. \\
}

\date{\today}

\maketitle


\begin{abstract}

We discuss the non--relativistic multichannel quark model and
describe the techniques developed to solve the resulting
equations.
We then investigate some simple solutions to demonstrate how the
model unifies meson--meson scattering with meson spectroscopy,
thereby greatly extending the domain of applicability of the
\naive quark model.
In the limits of narrow resonance widths and no quark exchange,
it reproduces the standard quark model spectroscopy and
Breit--Wigner phase description.
Outside those limits $s$--channel resonance masses are lowered
by their two--meson couplings, the line--shapes of wide
resonances are significantly altered, and the equivalent
Breit--Wigner masses and widths show an energy dependence.
Because meson--meson interactions are due to coherent
$s$--channel resonance production and $t$--channel quark
exchange (though other interactions can readily be added),
the multichannel equations model experimental resonance
production and decay in a way that the usual eigenvalue
equations cannot.

\end{abstract}


\section{INTRODUCTION}
\label{sec:intro}

The quark model attempts to represent the main features of
QCD, the theory of quarks and gluons,
and has been successfully used to calculate hadron spectroscopy
and decays
\cite{qmrefs}.
However, after decades of study, and numerous extensions, a
multitude of discrepancies still remain between theory and
experiment, particularly in the light meson sector.

In this paper we construct and solve a multichannel quark
model, which we apply to meson--meson scattering.
We consider only low energy scattering processes, which are
resonably soft
for all mesons except the $\pi$ at the high end of our
energy range, so we use a non--relativistic formulation.
We motivate the model phenomenologically by considering strong
decay processes, and add terms representing these processes to
the \naive quark model.
Numerical solutions, which are described in the Appendicies,
yield the $S$--matrix parameters for the free
two--meson states and the masses, widths, and two--meson decay
predictions for the $s$--channel meson resonances.
The standard quark model spectroscopy and the Breit--Wigner
phase shift
\cite{BreitWigner}
are reproduced, in the limits of narrow resonance width and
vanishing quark exchange.
Outside those limits resonance properties are found to depend
significantly on two--meson couplings,
which leads to mass shifts,
energy dependent Breit--Wigner parameters, and
two--meson bound states.

The paper is organized as follows:
We first review spectroscopy in the \naive quark model.
Next we develop the multichannel equation for the simple case of
the elastic scattering of two stable mesons, $A$ and $B$, through
an intermediate resonance $R$.
Some qualitative behaviours of the solutions are explored,
new effects are highlighted, and
likely avenues of further investigations are noted.
We then review the quark exchange process and incorporate it
into the multichannel model.
Finally we apply the model to inelastic multichannel scattering.

This model has been discussed previously in the square well
approximation
\cite{jw91,jw93}
and, in a less complete form in
\cite{jw_cebaf}.
See also
\cite{evb}.

The fortran source code that has been developed to find the
numerical
solutions to the multichannel equations, and rudimentary
intructions for running it, are available over the World Wide
Web
\cite{fortran}.


\section{THE NA\" IVE QUARK MODEL}
\label{sec:nrqm}

One method of predicting the meson spectrum from the \naive
quark model is to solve the eigenvalue problem
\beq
H_\Rqq ({\bf r}_\qq) \, \psi_\Rqq ({\bf r}_\qq)=
m_\R \, \psi_\Rqq ({\bf r}_\qq)
\eeq
for each $\qq$ meson resonance $R$.
Here the $\qq$ Hamiltonian $H_\Rqq ({\bf r}_\qq)$
defines all the properties of $R$.
In the \naive quark model we use a phenomenological $\qq$
central potential $V_\Rqq(\rqq)$, so that equation~(1) becomes
\beq
\label{eq:h1}
H_\Rqq ({\bf r}_\qq) =
\left[ -{\hbar^2 \over {2 m}} {\bf \nabla\cdot\nabla }
\right]
\, + m_q \, + \, m_{\bar q} \, + V_\Rqq \, (\rqq) \, ,
\eeq
where $V_\Rqq (\rqq)$ takes the simple form
\beqa
\label{eq:vqqbar}
V_\Rqq \, (\rqq)=
& - &
\! \!
\Biggl[
\, {3 \over 4}\, b\, \rqq
\, -{\alpha_c \over \rqq}
\, -\alpha_g \, {\sigma_g^3 \over \pi^{\sss 3/2} }
\, e^{-\sigma_g^2 \rqq^2}
\, + {3 \over 4}\, C
\Biggr]
\, \FF
\nonumber \\ \noalign{\vskip 4 pt}
& - &
{8\pi \over 3} {\alpha_h \over m_q m_{\bar q}}
\, {\sigma_h^3 \over \pi^{\sss 3/2} }
\, e^{-\sigma_h^2 \rqq^2}
\, {\bf S}_q \cdot {\bf S}_{\bar q}
\, \FF \, .
\eeqa
The eight components of ${\bf F}_q$ (${\bf F}_{\bar q}$)
are defined by
$[{\bf F}_q]^a={1\over2}{\bf \lambda^a}$
($[{\bf F}_{\bar q}]^a=-{1\over2}(\lambda^a)^*$), for $a=1,8$,
where the $\lambda^a$ are the eight 3x3 SU(3) color matricies.
The ${\bf S}_q$ and ${\bf S}_{\bar q}$ are defined by
$[{\bf S}_q]^i=[{\bf S}_{\bar q}]^i={1\over2}\sigma^i$, for $i=1,3$,
where the $\sigma$ are the three 2x2 SU(2) spin matricies.
The expectation value of the $\FF$ factor reduces to a constant
factor of $-4/3$ for mesons,
while $\langle {\bf S}_q \cdot {\bf S}_{\bar q} \rangle =-3/4$ for
spin zero $\qq$ systems and +1/4 for spin 1 systems
\cite{qmrefs}.

The linear confinement term ($br_\qq$) is predicted by the
flux--tube model and by lattice QCD.
The $\alpha_c/\rqq$ term is a short--range Coulomb piece that
dominates the interaction when $r_\qq \ll $~1~fm.
It can also be represented by a short
range gaussian potential, which
approximates the mid-- and long--range behaviour of the $1/\rqq$
interaction, but is finite at $\rqq = 0$.
$C$ is a constant background potential.
The last term is the short--range color magnetic hyperfine
interaction.
It increases (decreases) the energy of systems whose color magnetic
moments are aligned (anti--aligned).
We ignore spin--orbit coupling terms, which normally lead to
mass splitting of order 10~MeV between, for example, the
$^3P_0$, $^3P_1$, and $^3P_2$ states.
In
Section~\ref{sec:schannel}
we will show, for the $\ell_\qq=1$, $s_\qq=1$, $K_0$,
$K_1$, and $K_2$ states specifically, that different couplings
and different external two--meson states shift the
underlying $\qq$ resonance masses by 10s of~MeV.

If we use separation of variables to write the $\qq$
wavefunction as
\beq
\psi_\Rqq({\bf r}_\qq)
= {u_\Rqq(\rqq)}\, {1\over r_\qq} \, Y_\Rqq (\theta,\phi)
\eeq
then $u_\Rqq(\rqq)$ obeys the radial equation
\cite{merzbacher}
\beq
\label{eq:schr1}
H_\Rqq (\rqq) \, u_\Rqq (\rqq)= m_\R \, u_\Rqq (\rqq)
\eeq
where $H_\Rqq(\rqq)$ is the radial part of the $\qq$ Hamiltonian
given by
\beq
\label{eq:hqqbar}
H_\Rqq (r_\qq) =
-{\hbar^2\over2\,\mu_{\qq}}{{\rm d}^2 \over {\rm d}\rqq \,^2 }
+{\hbar^2\over2\,\mu_{\qq}}\, {\ell_\qq(\ell_\qq+1) \over \rqq^2}
\, + m_q \, + \, m_{\bar q} \, + V_\Rqq \, (\rqq).
\eeq
In (\ref{eq:hqqbar})
$\mu_\qq$ is the reduced $\qq$ mass, given by
$\mu_\qq=2m_{\sss q}m_{\sss \bar q}/(m_{\sss q}+m_{\sss \bar q})$.
The second term is the centrifugal barrier term, which comes
from the three dimensional kinetic energy operator.
Since we limit ourselves to this one--dimensional equation, we
choose to view the barrier term as part of the potential, rather
than part of the kinetic energy.

Solutions of
equation~(\ref{eq:schr1})
yield meson masses $m_\R$ and meson wavefunctions $u_\Rqq(\rqq)$
which, here, depend on quark flavor, $\qq$ separation $\rqq$,
spin $s_\qq$, and orbital angular momentum $\ell_\qq$.
The light quark sector of this
Hamiltonian has ten parameters,
$m_u$, $m_d$, $m_s$, $b$, $\alpha_c$, $\alpha_g$, $\sigma_g$,
$C$, $\alpha_h$, and $\sigma_h$,
which are fixed so the model reproduces a small subset of the
available data.
Predictions for the masses of all other light mesons follow.

As we are using constituent, rather than current quark masses,
we impose the constraint $m_u\!=\!m_d$, reducing the number of
parameters to nine.
We also choose $\alpha_g=0$, so the value of $\sigma_g$
becomes irrelevant, and we have seven free parameters.

In this paper we use
\hbox{$m_u = m_d = 0.375$~GeV,}
\hbox{$m_s = 0.650$~GeV,}
\hbox{$b = 0.782$~GeV/fm,}
\hbox{$\alpha_c = 0.857$,}
\hbox{$C=-0.581$~GeV,}
\hbox{$\alpha_h = 0.840$},
and
\hbox{$\sigma_h = 0.700$~GeV},
which leads to the meson masses quoted below.

In Section~\ref{sec:schannel} we will add two more parameters that
couple $\qq$ resonances and two--meson states, and the scattering
properties of two--meson systems can then be determined.

The wavefunctions $u_\Rqq(\rqq)$ solving
equation~(\ref{eq:schr1})
are found numerically by finite difference methods using the
Noumerov technique
\cite{noumerov}.
The details are provided in Appendix~\ref{apndx:noumerov}.
In the large $r$ region, these numerical solutions
are, generally, linear combinations of an exponentially decaying and an
exponentially growing term.
The relative amplitude of each term depends on the trial energy
$E_\R$.
The physical solutions are defined by the additional constraint that
$u_\R (\rqq) \to 0$ as ${\rqq\to\infty}$,
which requires that the amplitude of the exponentially growing
factor vanish, and defines the eigenenergies ${m_\R}$.
Details are given in Appendix~\ref{apndx:asymptotic}.

In
FIG.~1. we plot the $\ell_\qq=1$, $s_\qq=1$, $K_{J=0,1,2}$ radial
wavefunctions at trial energies of
$E_1=1.4299235012$~GeV and
$E_2=1.4299235014$~GeV.
(We have no spin--orbit term in the Hamiltonian, so the $K_J$'s
are degenerate at this point.)
The lower energy solution goes to $+\infty$ as $\rqq\to
\infty$ while the other solution, which is only 0.2 eV greater in
energy, goes to $-\infty$ as $\rqq\to \infty$!
The eigenenergy lies somewhere in between these two values, so
we can clearly use this proceedure to determine the
eigenenergies of
(\ref{eq:schr1})
to an irrelevantly high accuracy.

As a verification of this technique we tested the iterating
program on the toy problem of two quarks confined by an harmonic
oscillator potential.
We found that the trial energies differ from the exact analytic
eigenenergies by less than one part in $10^{11}$,
and the numerical wavefunctions differ from the exact gaussian
solutions by less than one part in $10^8$ out to distances of
8~fm!
The harmonic oscillator is a particularly simple potential
however, and we must always check the solutions to make sure
they are independent of the various iteration parameters
introduced to numerically solve equation
(\ref{eq:schr1}).

In FIG.~2. we show the wavefunction of the first radial
excitation of the $K_J$ at energies of
$E_1=2.047471763$~GeV and
$E_2=2.047471768$~GeV.
This energy resolution of 5 eV is coarser than above simply to
make the wavefunctions diverge a distances of less than 6.0~fm.
(As the trial energy approaches the eigenenergy the point of
divergence of the trial wavefunction moves further from the
origin.)
The second radial excitation is at 2.5651608366~GeV, and the higher
radial excitations are easily found this way.

With the Hamiltonian parameters above,
this approach yields meson masses (in~GeV) of
$m_\pi = 0.133 $,
$m_K = 0.494 $,
$m_\rho = 0.775 $,
$m_{K^*} = 0.913 $,
$m_\phi = 1.019 $,
$m_{b_1} = 1.230 $,
$m_{K_1} = 1.389 $,
$m_{h'_1} = 1.512 $,
$m_{a_J} = 1.280 $,
$m_{K_J} = 1.430 $, and
$m_{f'_J} = 1.550 $,
with J=1, 2, and 3, in good agreement with experiment.


\section{s--CHANNEL MESON RESONANCE PRODUCTION}
\label{sec:schannel}

When solving any $\qq$ eigenvalue problem one implicitly
assumes that resulting mesons have a unique energy, and are
therefore stable.
However, most hadrons are hadronically unstable,
with typical widths of 200~MeV.
The Heisenberg uncertainty principal gives them a
lifetime of
\beq
\tau
\approx {\hbar \over \Delta E}
\approx 10^{-15} {\rm fm}
\approx 3 \,{\rm x}\, 10^{-24} {\rm sec}\, ,
\eeq
or as long (short) as it takes light to cross a meson.
The zero width approximation intrinsic to the eigenvalue
formulation is, therefore, unrealistic for all the hadronically
decaying hadrons.

When we try to extract the spectroscopy of wide and
interfering hadrons, glueballs, and hybrids from experimental
data, as a means of gathering information about the underlying
quark--quark interactions, and to test theories of QCD, we find
the predictions of
equation~(\ref{eq:schr1}) inconsistent with data.
There are results, such as the presence of two $\rho^\prime$
states, which have remained unexplained after years of intense
research and many extensions of~(\ref{eq:schr1}).

The multichannel model we propose scatters two--meson
states into two--meson
states through $s$--channel resonance production and
$t$--channel quark exchange.
The relative wavefunction of two--meson system, and the
wavefunction of $s$--channel resonance, are found.
{}From these the $S$--matrix and the resonance parameters are
deduced.

Consider $s$--channel meson resonance formation.
As an idealized gedanken scattering experiment one might think
of two mesons, $A$ and $B$, coalescing via $\qq$ annihilation to
form a single meson resonance $R$.
(We will include the $t$--channel quark exchange effects below.)
Some time later $R$ decays, via $\qq$ creation, into a final
two--meson system $CD$, where $CD$ may or may not represent the
same mesons as $AB$.
This is depicted in
FIG.~3.

For the moment we restrict our discussion to the elastic
scattering process
$AB \to R \to AB$.
Consider the initial state $AB$ in
FIG.~3.
The \schrod equation, which determines the
relative $AB$ wavefunction $u_\AB(r_\AB)$, is
\beq
\label{eq:H_AB}
H_\AB (\rAB) \, u_\AB (\rAB)
\, = E_\AB
\, u_\AB (\rAB) \, ,
\eeq
where the radial part of the two--meson Hamiltonian $H_\AB(\rAB)$ is
given by
\beq
\label{eq:H_AB1}
H_\AB (\rAB) =
-{\hbar^2 \over 2\, \mu_{\AB} } \left(
\, {{\rm d}^2 \over {\rm d}\rAB \,^2 }
\, -{\ell_\AB(\ell_\AB+1) \over \rAB^2}
\right)
\, + V_\ABAB (\rAB)
\, + m_\A
\, + m_\B \,.
\eeq
We have included the meson masses, $m_\A$ and $m_\B$,
in the Hamiltonian so that the
energy $E_\AB$ represents the total energy of the $AB$
system in its center of momentum frame.
The potential $V_\ABAB$ will represent other possible
non--resonant $AB$ interactions, such as quark exchange shown in
FIG.~4.,
and will be discussed in
Section~\ref{sec:qx}.
For the moment we take $V_\ABAB=0$.

Equation~(\ref{eq:H_AB})
says that the $AB$ energy operator, $H_\AB$, acting on the $AB$
probability amplitude, $u_\AB(\rAB)$,
equals the availiable energy $E_\AB$ times $u_\AB(\rAB)$.
When one looks at the transition $AB$ to $R$ in
FIG.~3.,
however, one sees that $R$ removes
energy from $AB$.
This suggests modifying equation~(\ref{eq:H_AB}) by removing a
separation dependent energy factor,
$V_\ABtoR (\rAB) \, u_\qq (\rqq)$, from the right hand side of
(\ref{eq:H_AB}), leading to
\beq
\label{eq:H_ABR}
H_\AB \, u_\AB (\rAB)
\  = E_\AB
\, u_\AB (\rAB) \\
\, - V_\ABtoR (\rAB)
\, u_\qq (\rqq).
\eeq

A standard model for $\qq$ annihilation couples the
$\qq$ state to
the vacuum, which requires that
the annihilating $\qq$ pair be in an $\ell_\qq =1$,
$s_\qq =1$, $^3P_0$ color singlet state
\cite{yao,barnes96}.
It is, however, possible that some annihilation verticies are
dominated by the process $\qq \to g$ (where this $g$ is a
gluon).
This would require that the annihilating $\qq$ pair have the
$^3S_1$ quantum numbers of the gluon, constraining them to an
$\ell_\qq =0$, $s_\qq =1$, color octet state, and making $R$
a hybrid ($\qq g$) meson.
The model we are exploring here can be adapted to quantitatively
study this fundamental question, though we do not do so here.

In this paper we choose, as an ansatz for the creation and
annihilation verticies, the transition potential
\beq
\label{eq:V_ABR}
V_\ABtoR (\rAB) =
g\,
C^f_\ABR \,
C^s_\ABR \,
\left[{\hbar\over r_{\sss 0}}\right] \,
N_{\ell_\AB} \,
\left[{r_\AB\over r_{\sss 0}}\right]^{\ell_\AB} \,
{\rm e}^{-(r_\AB /r_{\sss 0})^2}
\eeq
where $g$ is an unknown universal constant (with dimensions of
energy) to be
determined from some subset of experimental data.
$C^s_\ABR$ and $C^f_\ABR$
are, respectively, the spin and flavor Clebsch--Gordon
factors relating the overlap of the $AB$ and $R$ spin and flavor
states, and depend on the structure of the $\qq$ creation and
annihilation verticies.
We generally take $C^s_\ABR = C^f_\ABR = 1$
here, an approximation that needs to be removed before
quantitative predictions can be realized.
$\ell_\AB$ indexes the angular momentum of $AB$.
The second unknown parameter in (\ref{eq:V_ABR})
is $r_{\sss 0}$, which sets the
range of the interaction.
It must be determined from fits to experiment but is expected,
from \naive arguments, to be approximately 0.5~fm.
In this paper we use $r_{\sss 0}=3.0$~GeV$^{-1}\approx 0.6$~fm.
$N_{\ell_\AB}$ is a normalization constant defined by the
constraint
\beq
N_{\ell_\AB}\int_0^\infty
\left[{r_\AB\over r_{\sss 0}}\right]^{(\ell_\AB+2)} \,
{\rm e}^{-(r_\AB /r_{\sss 0})^2} \,
{dr_\AB\over r_{\sss 0}}
=1 \,.
\eeq

Other forms of the transistion potential $V_\ABR (\rAB)$ are
obviously likely
\cite{barnes96},
and the exploration of these possibilities is an important
application of the multichannel model.

Notice that both $\rAB$ and $\rqq$ appear in
equation~(\ref{eq:H_ABR}),
so we must relate them.
An examination of the $\qq\,\qq$ geometry
(FIG.~5.)
an instant before the transition into the $\qq$ state
shows, in the SU(3) limit where $m_u=m_d=m_s$,
that $r_\R\equiv\rqq=2\,\rAB$.
This relationship is easy to implement in
(\ref{eq:H_ABR}).
(See Appendix~\ref{apndx:2cse}.)

The $\qq$ resonance $R$ in
FIG.~3.
was described by equation~(\ref{eq:schr1}).
{}From
FIG.~3.,
however, we see that it must be modified to allow
for the transfer of the energy from $R$ into the $AB$ system,
leading to
\beq
\label{eq:H_RAB}
H_\Rqq \, u_\Rqq (\rqq)
= E_\AB
\, u_\Rqq (\rqq) \\
- V_\RAB (\rAB)
\, u_\AB (\rAB) \, .
\eeq
By time reversal invariance we require that
$V_\ABtoR = V_\RAB \equiv V_\ABR$.


\section{The Two--Channel Model}
\label{sec:twochannel}

Equations
(\ref{eq:H_ABR}) and
(\ref{eq:H_RAB})
can be combined into matrix form:
\beq
\label{eq:ccm}
\left[ \,
\begin{array}{cc}
H_\AB (\rAB) & V_\ABR (\rAB) \\
V_\ABR (\rAB) & H_\Rqq (\rqq)
\end{array}
\, \right]
\, \left[ \,
\begin{array}{c}
u_\AB (\rAB) \\ u_\Rqq (\rqq)
\end{array}
\, \right]
= E_\AB
\, \left[ \,
\begin{array}{c}
u_\AB (\rAB) \\ u_\Rqq (\rqq)
\end{array}
\, \right] \, ,
\eeq
where $E_\AB$ is the energy available in the $AB$ center of
momentum system and takes on values greater than $m_\A+m_\B$
\cite{jw91,jw93}.
{\it Equation~(\ref{eq:ccm}) is the 2--channel realization of the
multichannel model.}

The $\qq$ Hamiltonian is embedded in (\ref{eq:ccm}), but
{\it in this matrix formulation we deduce the properties of
the resonance $R$ due to both its $\qq$ interactions and to its $AB$
couplings}.
Thus, finite lifetime effects are included, {\it a priori},
in the properties of the $\qq$ resonances.
Solutions of the eigenvalue equation (\ref{eq:schr1})
can be recovered from (\ref{eq:ccm}) by allowing the
off--diagonal potentials
$V_\ABR$ to become vanishingly small, effectively decoupling $R$
from $AB$.
We demonstrate this below.

All we need do is solve
equation~(\ref{eq:ccm})
for the wavefunctions $u_\AB (\rAB)$ and $u_\Rqq (\rqq)$!
This will provide us all the information discernable
about the $AB\lra R$ system.
Describing the solutions of equation~(\ref{eq:ccm}) is left to
Appendicies A, B, and C.

{}From the solutions we extract the energy dependent $AB$
phase shift, $\delta^{2ch}(E_\AB)$,
(see equation~(\ref{eq:delta2})) due to the presence of the
resonance $R$, and from $\delta^{2ch}(E_\AB)$, the mass and
width of $R$.
We define the width as the energy difference between the energy at
$\delta^{2ch}=135^\circ$ and
$\delta^{2ch}=45^\circ$,
and the resonance mass as the energy at which 
$\delta^{2ch}=90^\circ$.

A standard description of the phase shift for the process
$AB \to R \to AB$
is the non--relativistic Breit--Wigner given by
\beq
\label{eq:delta_bw_AB}
\delta^{\BW}_\AB
={\rm arctan}\Biggl({{1\over2}\,\Gamma_\R \over \mR -E_\AB}\Biggr)
\, ,
\eeq
where $\mR$ and $\Gamma_\R$ are, respectively,
the Breit--Wigner mass and width of the resonance.
Choosing the quark exchange potential
$V_\ABAB \! =\! 0$
in equation~(\ref{eq:H_AB1}), and therefore (\ref{eq:ccm}),
leaves $s$--channel $R$ production as
the only interaction mechanism available to $A$ and $B$.
Therefore, the resulting two--channel phase shift, given by
equation~(\ref{eq:delta2}) in Appendix~\ref{apndx:2cse}, should
closely approximate the Breit--Wigner phase shift if
the multichannel model is to be viable.

In
FIG.~6. we plot the two--channel and the
Breit--Wigner phase shifts for $s$--wave $K\pi$
elastic scattering through the $K_0$ scalar meson.
Their excellent agreement establishes this essential and
non--trivial connection between these two approaches.

In solving this toy problem we picked a
$V_\ABR$ coupling strength $g$ that gave the $K_0$ a width of
only 55~MeV, rather than the physical value of $287\pm10$~MeV
\cite{pdg}.
As we shall see below, phase shifts found from scattering
through wide resonances deviate from the Breit--Wigner lineshape.
Also note that the $K_0$ phase shift now passes through
$90^\circ$ at 1.412~GeV, down from
the eigenenergy value of 1.430~GeV quoted earlier.
{\it This mass shift is due entirely to the coupling
between the
$K_0$ and its production and decay $K\pi$ channel}, and is a
general feature of multichannel systems.

Another required, and non--trivial, connection between this
model and the \naive quark model is that they share spectroscopies
in the limits of narrow resonances and no quark exchange.
In
FIG.~7. we plot the $K\pi$ phase shifts for different values of $g$.
At the smallest value of $g$ the $K_0$ resonant state
has a mass of 1.429~GeV and a width of 6~MeV,
in excellent agreement with the zero width eigenvalue of
1.430~GeV, demonstrating that this correspondence requirement is
satisfied.
(Were we to make the $K_0$ still narrower, the scattering energy
would get still closer to the eigenenergy.)

Also shown in
FIG.~7. are phase shifts
at larger values of $g$.
As the lifetime of the resonant state $R$ decreases
its' mass also decreases and the Breit--Wigner lineshape is
lost.
Above a critical value of $g$, in this case at about $g=1.40$~GeV, the
$K_0$ mass falls below the $K\pi$ threshold, thereby making the
$K_0$ a linear combination of $\qq$ and bound $s$--wave $K\pi$
components, and stable against strong decay.
Again, this is entirely a multichannel effect.

In
FIG.~8. we plot sin$^2(\delta^{2ch}_{K\pi})$ for $K\pi \lra K_0$
$s$--wave elastic scattering,
with $g$ chosen to generate a $K_0$ width of 280~MeV,
within the errors of its experimental value.
We now find the mass of the ground state $K_0$ is
1.355~GeV, a full 75~MeV below the eigenenergy of the state in
the zero width approximation.
Obviously the Hamiltonian parameters need to be modified for the
multichannel model.
The eigenvalue approach can be thought of as generating the
masses of ``bare" $\qq$ states; while the multichannel approach
``dresses" the bare states with connected two--meson channels,
making them lighter.
The data used to fit the bare Hamiltonian parameters has necessarily
been the dressed $\qq$ spectra, that is all we have access to
experimentally.
This mismatch needs to be recitified; the dressed Hamiltonian
parameters must be tuned to the dressed data.

In
FIG.~8. we take the $K\pi$ scattering energy all the
way up to 3.0~GeV, and see the first two radial excitations of the
$K_0$ at 2.020 and 2.553~GeV,
again showing the close connection between scattering and
spectroscopy.
The masses of the first radial and the second radial states
are, respectively, only 27 and 12~MeV below their
eigenenergies.

It is interesting to explore the behaviour of the $K_0$
resonance produced in $s$--wave $K\pi$ scattering.
We made a movie of the energy evolution of the scattering
and $\qq$ wavefunctions but were unable to draw any conclusions
from it
\cite{jw_js}.
In
FIG.~9. we plot the energy dependent
probability of finding some $\qq$ state with $K_0$ quantum
numbers at the core of the
scattering system.
This probability is given by
\beq
P_{K_0}(E_{K\pi})=
\int_0^\infty | u_{K_{\sss 0}}^\ast u_{K_{\sss 0}} |^2\, dr
\, ,
\eeq
where the $u_{K_0}$ normalization is fixed by setting the
ampltitude of the external $K\pi$ wavefunction equal to unity
(see Appendix~\ref{apndx:2cse}).
We observe the surprising result that this probability
never vanishes.
In a beam--beam experiment, at phase shifts of integer
multiples of $\pi$ radians,
we find vanishing cross sections and therefore expect that no
$s$--channel $\qq$ resonance is produced.
We have, however, constructed a model in which the incident
state is a shell of inward falling $K\pi$ probability amplitude,
and not two colliding beams.
In this picture it is easy to visualize an outward moving shell
being phase shifted from an inward moving shell by integer
multiples of $\pi$ radians due to transistions to central
$\qq$ states.

We can fix the $J$ of the $K_J$ state by fixing the vertex
coupling and the external states, thus breaking the $K_0$,
$K_1$, $K_2$ degeneracy which is present in our
eigenstate Hamiltonian.
In
FIG.~10. we show the line shapes for the three processes
$(K\pi)_s \lra K_0$,
$(K^\ast\pi)_s \lra K_1$, and
$(K\pi)_d \lra K_2$ scattering, where the subscript on the
two--meson states specifies their relative orbital quantum
number.
We see that the $K_J$ now appear at different masses.
The $K_2$--$K_0$ mass splitting of 22~MeV arises from their
$d$--wave and $s$--wave $K\pi$ couplings respectively.
The $K_0$--$K_1$ mass
difference of 17~MeV arises from $K\pi$ versus $K^\ast
\pi$ couplings.
These splittings are of the same magnitude as the spin--orbit
splittings in the eigenvalue quark model
\cite{qmrefs}.

The above discussion also suggests that, when two resonances
with the same quantum numbers are found at nearby masses in
different channels, the correct interpretation is likely that
they reveal the same underlying state, and that the mass
differences are due entirely due to different couplings.
Testing this conjecture in all resonance channels is an
essential step in proving the veracity of the multichannel
approach.
We give an example of this effect in the $K\pi,\ K\eta,\
K\eta^\prime,\ K_0$ system below.

It is important to emphasize that we are currently only able to
demonstrate the qualitative behavior of the multichannel
solutions.
Specific result will require the implementation of
more accurate models for the $AB \lra R$ verticies,
the inclusion of all two--meson channels in each sector
(which will require the multichannel model discussed below),
and a re--tuning of the Hamiltonian parameters to some small
subset of the low--energy data.
While quantitative predictions are not yet available the
essential points are clear, obvious, and intriguing;
$s$--channel meson resonance properties are significantly
modified by their two--meson couplings.
This multichannel model is an excellent vehicle to begin
this exploration.


\section{The Energy Dependence of the Breit--Wigner Parameters}
\label{sec:bw_ed}

Equation (\ref{eq:delta2}) gives us the energy dependent phase
shift in the two--channel model.
We can use the resulting curve, $\delta^{{\sss 2ch}}(E)$,
to find the
energy dependence of the Breit--Wigner mass and width parameters
of the resonance $R$.

Suppose we find
$\delta^{{\sss 2ch}}(E_1)$ and
$\delta^{{\sss 2ch}}(E_2)$
at two neighboring energies $E_1$ and $E_2$.
Equating $\delta^{{\sss 2ch}}$, from equation~(\ref{eq:delta2}),
and $\delta^\BW$, from equation~(\ref{eq:delta_bw_AB}),
at these two energies leads to
\beq
\zeta_1\equiv
\tan(\delta^{{\sss 2ch}}(E_1))={{1\over2}\,\Gamma_\R \over \mR -E_1}
\hskip 0.5 true cm
{\rm and}
\hskip 0.5 true cm
\zeta_2\equiv
\tan(\delta^{{\sss 2ch}}(E_2))={{1\over2}\,\Gamma_\R \over \mR -E_2}
\eeq
from which it follows that
\beq
\mR={E_2\,\zeta_2 - E_1\,\zeta_1 \over \zeta_2 - \zeta_1}
\eeq
and
\beq
\Gamma_{\R} = 2 \, (\mR-E_1) \, \zeta_1 \, .
\eeq

Using these energy dependent parameters in the
Breit--Wigner phase shift equation would reproduce the
two--channel phase shift for arbitrarily wide resonances.
The experimental implications are simply that the
amplitudes used to represent intermediate isobars
must allow for variable isobar masses and widths in different
regions of the Dalitz Plot.
In
FIG.~11. and
FIG.~12. we plot $\mR(E)$ and $\Gamma_\R(E)$
for the two--channel $(K\pi)_s \lra K_0$ system as an example of
their behaviour.
Again, specific predictions are required on a channel by
channel basis.

The idea of a variable resonance width has been discussed from
the point of view of final state interactions
in, for example, the reaction $\iota\to\eta\pi\pi/\eta K\bar K$
\cite{iota}.
Here we see that the changing mass and width of the
resonance states need not depend on final state interactions
{\it per se}, they arise simply from resonance couplings to final
states.

\goodbreak


\section{QUARK EXCHANGE EFFECTS}
\label{sec:qx}

We now consider the possibility of mediating to two--meson
scattering by $t$--channel quark exchange effects.

We omit $t$--channel meson exchange for three reasons
\cite{jw93}.
Meson exchange is topologically equivalent to $s$-channel
resonance production and therefore already, at some level,
included,
it is higer order in $\alpha_{\rm \sss strong}$ than $t$-channel
quark exchange, and
the range for meson exchange is of order
$\hbar/m_{\rm \sss meson}$, which is
less than 1~fm for all mesons except the $\pi$.
At these separations two hadrons, each $\approx \! 1$~fm in
diameter, have wavefunctions which overlap significantly,
and quark exchange seems highly probable.

Much effort has gone into explaining hadron scattering in terms of
meson exchange models, however, and it would be foolhardy to
dismiss this
possibility outright, especially at larger momentum transfers.
In fact, by time--reversing $A$ and $C$ in
FIG.~4. we
can see that $t$--channel quark exchange can lead to $\qq$
exchange diagrams, which may mock--up meson exchange,
\cite{jw93,jw_cebaf}.
Fortunately the multichannel model again offers an excellent
vehicle to study the inclusion of these terms; they will add to
the non--resonant potentials $V_\ABCD$, which are generalizations
of $V_\ABAB$ in equation~(\ref{eq:H_AB1}).

We were originally led to realize the importance of quark exchange
in a variational calculation of the $\qqqq$ ground state
wavefunction, $\psi_{0\, \qqqq}$, in the $J^P\! =\! 0^+$ sector
\cite{jw93,jwni}.
$\psi_{0\, \qqqq}$
required symmetry terms, embodying quark exchange,
and the short--range hyperfine interaction, to represent interacting
mesons.
Without either feature $\psi_{0\, \qqqq}$
represented non--interacting mesons.

The $\qqqq$ Hamiltonian used to obtain $\psi_{0\, \qqqq}$ is
just a generalization of the $\qq$ Hamiltonian of
equation~(\ref{eq:hqqbar}),
but has added to it a weak color independent quadratic potential,
\beq
\label{eq:v_eps}
V_\epsilon
\equiv
\sum_{i\ne j} {_1 \over^2} \epsilon r_{ij}^2 \, ,
\eeq
between all constituent pairs $ij$ that prevents repulsive and
non--binding two--meson systems from drifting beyond their
interaction range.
We interpreted the wavefunction
\beq
\label{eq:psi_extracted}
\psi_\AB({r}_\AB)
=
\langle \psi_\A \, \psi_\B \vert
\, \delta ( {r}_\A - {r}_\B + {r}_\AB ) \vert
\psi_{0\, \qqqq} \rangle
\eeq
as the amplitude for finding meson
$A$ at ${r}_\A$ with wavefunction $\psi_\A$,
and meson
$B$ at ${r}_\B$ with wavefunction $\psi_\B$,
in $\psi_{0\, \qqqq}$ and separated by distance
${r}_\AB$.
Inverting the radial Schr\"{o}dinger equation for
$\psi_\AB$
allowed us to extract an effective central $AB$ interaction potential
$V_\ABAB$:
\beq
\label{eq:v_extracted}
V_\ABAB (r_\AB) =
{\hbar^2 \over 2\, \mu_\AB }
\,
\biggl[
{ {\rm d}^2
\over
{\rm d} r_\AB^{\, 2}}
\, \psi_\AB
(r_\AB)
\biggr]
\psi_\AB^{-1}
(r_\AB)
\, + {\rm E}_\AB
\, - 4\, {_1 \over^2} \epsilon \,
r_\AB^2
\eeq
where $E_\AB$ is the binding energy of the interacting mesons.
These led to Gaussian--like potentials which can be used
in a Schr\"{o}dinger equation, either
directly or as ``equivalent'' square well potentials,
to find, for example, the phase
shifts for I=3/2 $K\pi$ and I=2 $\pi\pi$ scattering
\cite{jw_cebaf}.
These two processes have no $s$--channel resonances
to mediate their interactions, so they must go through either
$t$--channel quark exchange, $t$--channel meson exchange, or
some combination of the two.
The phase shifts resulting from the $t$--channel quark exchange
potentials compare favorably with experiment
\cite{estabrooks,wagner}.
SU(3) relations also allow quark exchange potentials to be
deduced for off--diagonal processes (such as $\pi\pi \rightarrow
\eta\eta$) from the SU(3) diagonal potentials
\cite{jw93,jwni}.

Based on the successes of this technique Barnes {\it et\,al.}
\cite{barnes,swanson}
developed a more widely applicable Born approximation technique
for extracting potentials based on one gluon exchange
followed by quark (or antiquark) exchange, with interactions
normally mediated by the short--range hyperfine interaction.
For the particular case of meson--meson scattering Swanson
\cite{swanson}
has shown that the effective intermeson potentials can be
modelled accurately by the multi-Gaussian form
\beq
\label{eq:v_exchange}
V_{{\sss \! AB\rightarrow CD}} (r_\AB)
=\sum_i \, a_i \, {\rm e}^{-{1\over 2}(r_{\! AB}/b_i)^2} \, ,
\eeq
where the $a_i$ and $b_i$ depend on the particular reaction
$AB \rightarrow CD$.
Agreement between these two theoretical approaches and
experiment is encouraging, in those sectors where comparisons
have been made
\cite{estabrooks,wagner,barnes}.

We again choose $K\pi$ scattering through
the $K_0$ scalar resonance to demonstrate the effect of these
exchange potentials.
First note that the $K\pi\lra K_0$ system is also coupled to
the $K\eta$ and $K\eta^\prime$ systems, so this is a four--channel
problem containing three two--meson channels and one $\qq$ channel.
Techniques for solving these multichannel equations are given in
Appendix~\ref{apndx:mcse}.

In
FIG.~13. we plot LASS data for $I=1/2\ K\pi$
scattering
\cite{lass}
with (a) $K\pi$ $s$--wave phase shifts due to
two--channel
$K\pi \lra K_0$ scattering
with no quark exchange potentials,
(b) $K\pi \lra K\eta \lra K\eta^\prime \lra K_0$ scattering
with only $s$--channel couplings to the $K_0$,
(c) $K\pi \lra K\eta \lra K\eta^\prime$ scattering
with only quark exchange potentials, and
(d) $K\pi \lra K\eta \lra K\eta^\prime \lra K_0$ scattering
with quark exchange potentials.

In this system the ratio of the relative strengths of the
annihilation couplings, which we can write as
$g_{K\pi\to K_0}:g_{K\eta\to K_0}:g_{K\eta^\prime\to K_0}$
where these $g$'s include the Clebsch factors in
equation~(\ref{eq:V_ABR}), are
${\sqrt{6}/\mu_{K\pi}}:
{(1-\sqrt{2})/\mu_{K\eta}}:
{(1+\sqrt{2})/\mu_{K\eta^\prime}}$
\cite{jw93}.
The $t$--channel potentials are taken from the variational
calculation reviewed above.

The two--channel phase (a) undershoots the low energy data but fits
it in the mid--ranges.
The phase shift due to only $s$--channel resonance coupling
to the three two--meson states (b) undershoots the
data below 1.2~GeV and overshoots it above that energy.
The $t$--channel quark exchange phase shift (c) looks like an
``effective range polynomial background'' term,
suggesting that {\it the need to add a non--resonant
background term when carrying out experimental amplitude
analyses is actually a reflection of underlying $t$--channel
quark exchange dynamics}.
The full phase shift (d), including both $s$-- and $t$--channel
processes, is the only curve that fits the low--energy data,
though above 1.2~GeV it overshoots the data.

Recall that we are using ``bare" Hamiltonian parameters, and a
very \naive $\qq$ annihilation and creation vertex model with
the range chosen arbitrarily at $r_{\sss 0}=3.0$ GeV$^{-1}$, so
the deviation from the data is not a statement that the model
itself is inadequate, but rather a statement that more work is
needed to make these qualitative results quantitative.

As an example of resonance mass shifts
seen when an initial two--meson state scatters through a
multichannel system into different finals states, we consider
\beq
K\pi \to (K\pi\lra K\eta\lra K\eta^\prime\lra K_0) \to 
\cases{K\pi \cr K\eta \cr}
\eeq
In FIG. 14. we show
$\sin^2(\delta^{4ch}_{K\pi\to K\pi})$ and
$\sin^2(\delta^{4ch}_{K\pi\to K\eta})$ via these
four--channel intermediate states, using both
$s$--channel $K_0$ couplings and $t$--channel quark exchange.
The apparent $K_0$ mass differs by about 45~MeV between these two
reactions.


\section{CONCLUSIONS}
\label{sec:conclusions}

The multichannel quark model approach treats meson spectroscopy
and meson--meson scattering on an equal footing.
It reproduces the Breit--Wigner and meson spectroscopy results
for $s$--channel resonances in the limits of narrow resonance
widths and no quark exchange.
The effects of
strong resonance couplings to open decay channels
and quark exchange effects lead to significant new physics away
from these limits.

This approach must be applied to all of meson spectroscopy and
two--body scattering to fully test it as a model, and to extract
from it the maximum amount of information and understanding.

Of primary importance is the incorporation of realistic
$AB\lra R$ vertex potentials, which might even be energy dependent
functions that would be easily incorporated into the
multichannel code
\cite{fortran}.
Each two--meson sector must be considered, and all two--meson
and resonance states must be included.
Following this, the Hamiltonian parameters must be refit to some
low energy data.
At this point the model can be applied to the standard $\qq$
meson spectrum, where inconsistences abound,
as well as to studies of glueball signatures, hybrid signatures,
and different annihilation models.

This version of the multichannel model is limited by its
non--relativistic kinematics, its
treatment of external mesons as stable, and its inability to
model two--step processes such as 
\beq
AB \to 
\left(
\begin{array}{c}
(R_{\sss CD})\, E \\
(R_{\sss DE})\, C \\
(R_{\sss EC})\, D
\end{array}
\right)
\to CDE
\eeq
where
$A,\ B,\ C,\ D,$ and $E$ are
stable mesons, and
$R_{\sss CD} $,
$R_{\sss DE} $, and
$R_{\sss EC} $
are wide and interfering intermediate resonances.
Developing the model along these lines seems warrented.

The numbers of open avenues of exploration are numerous.
We have pointed out some of these along the way, and certainly
others remain to be discovered.

To facilitate the development of the model the fortran source
code written to solve the multichannel equations is available
over the World Wide Web
\cite{fortran}.
This code also includes some features that were not mentioned in
this paper.
For example:
\begin{enumerate}
\item[$\bullet$] It already contains a search routine (called
fitham.f) to refit the Hamiltonian parameters.
This routine calls the $\qq$ eigenfunction program (called
meson.f) to find the eigenenergies of the hadronically stable
hadrons, and the scattering program (called ccp.f, for coupled
channel program) to find the mass and width of hadronically
decaying resonances.
With access to this information it searches the Hamiltonian
parameter space for the best fit to the experimental data.
Constraints such as the value of the linear confinement
potential, or the expected quark masses, can easily be
implemented as part of the fit.

\item[$\bullet$] Both the meson eigenvalue program and the
scattering program have options of printing wavefunctions, so
the energy
evolution of the multichannel wavefunctions can be explored
\cite{jw_js}.
In particular, the contents of the scalar $f_0(975)$ and
$a_0(983)$ $K\bar K$ molecule states can be studied once the
dressed Hamiltonian parameters are found.
This will inevitably reveal that, for example, the $f_0(975)$ is
a mixture of
\hbox{$\pi\pi$},
\hbox{$K\bar K$},
\hbox{$\eta\eta$},
\hbox{$\eta\eta^\prime$},
\hbox{$\eta^\prime \eta^\prime$},
\hbox{$\sqrt{1\over2}(u\bar u+d\bar d)$}, and
\hbox{$s\bar s$},
and that the wavefunctions of each component will vary
dramatically in shape and relative strength even within the
narrow width of the $f_0$.

\item[$\bullet$] It contains a production model for studying
processes such as $J/\psi \to \phi\pi\pi$ and
$J/\psi \to \phi K\bar K$
\cite{production},
which likely proceed through
the intermediate $\phi\, f^\prime_0$ state.
The outgoing $\pi\pi$ or $K\bar K$ pairs result from the
creation of a central $s\bar s$ scalar state,
which subsequently bubbles out through the multichannel
scattering potentials into the final state.
For 
$J/\psi \to \omega\pi\pi$ and
$J/\psi \to \omega K\bar K$
one simply creates a central $\sqrt{1\over2}(u\bar u+d\bar d)$
scalar state, and so on.
These processes contain no incoming two--meson states.
Solutions are found by solving the
homogeneous and inhomogeneous multichannel equations,
and then forming the linear combination
of these solutions which have no incoming two--meson components.
The inhomogeneous multichannel equation is formed by adding a
$\delta(0)$ source term in one of the $\qq$ channels to the
right hand side of equation (D1).
This has the effect of
altering the boundary conditions on $\u(0)$ and $\u^\prime(0)$,
but leaving the rest of the equation unchanged.
The algebra is left as an exercise to the reader.

\item[$\bullet$] There is a sector for studying baryon--baryon
scattering,
in particular looking for the H--dibaryon and the deuteron as a
multichannel effect.
\end{enumerate}

The multichannel quark model is clearly in its infancy.
It is full of possibilities, and holds the promise of providing
many insights into hadron
spectroscopy and scattering dynamics.
It may prove to be an invaluable, perhaps even an essential, tool
for resolving the meson spectrum, understanding the $\qq$
annihilation verticies, and positively identifying the
long--sought glueball and hybrid states.
In the process it will be teaching us new ideas about hadron physics.

\vfill\eject

\section{Acknowledgements}

This work would not have been possible without the contributions
of Nathan Isgur and Teb Barnes, and was greatly advanced by
contributions, suggestions and questions from Gerry Mahan, Joe
Macek, Bill Dunwoodie, and Alex Dzierba.
I also thank the Experimental Group at the University of
Mississippi for providing computational platforms and software
support.
Finally, I am grateful to Leslie Cameron and Jon Dugger for their
constant and invaluable input.

\vfill\eject


\appendix
\section{The Noumerov Technique}
\label{apndx:noumerov}

\def\derI{^{\prime}}
\def\derII{^{\prime\prime}}
\def\derIII{^{\prime\prime\prime}}
\def\derIIII{^{\prime\prime\prime\prime}}
\def\derIIIIII{^{\prime\prime\prime\prime\prime\prime}}
\def\wf{\scriptstyle}

We introduce the Noumerov technique
\cite{noumerov}
used to numerically iterate the \schrod equation
\beq
\label{eq:a1}
-{\hbar^2 \over 2 \mu }\, u\derII(r)
+ V(r) \, u(r) = E \, u(r),
\eeq
with
\beq
u\derII (r) \equiv {d\over dr}\,{d\over dr}\, u(r)\, .
\eeq
By defining a modified ``potential'' function $W$ as
\beq
W(r)\equiv -{2\mu\over \hbar^2} \, (E-V(r))
\eeq
we can rewrite (\ref{eq:a1}) as
\beq
\label{eq:a4}
u\derII (r) = W(r) \, u(r) \, ,
\eeq

We make $r$ discrete, label the points
$r_i$, and set $\epsilon = r_{i+1}-r_i$.
Taylor expanding $u(r_{i+1})$ and $u(r_{i-1})$ leads to
\beqa
u(r_{i+1})
&=&
u(r_i)
+\epsilon\, u\derI(r_i)
+{\wf{1\over2}}\epsilon^2\, u\derII(r_i)
+{\wf{1\over6}}\epsilon^3\, u\derIII(r_i)
+{\wf{1\over24}}\epsilon^4\, u\derIIII(r_i)
+ \dots \\
\nonumber
u(r_{i-1})
&=&
u(r_i)
-\epsilon\, u\derI(r_i)
+{\wf{1\over2}}\epsilon^2\, u\derII(r_i)
-{\wf{1\over6}}\epsilon^3\, u\derIII(r_i)
+{\wf{1\over24}}\epsilon^4\, u\derIIII(r_i)
- \dots
\eeqa
and then to
\beq
\label{eq:aupm}
u(r_{i+1})+ u(r_{i-1})
=2\, u(r_i)
+ \epsilon^2\, u\derII(r_i)
+ {\wf{1\over12}}\epsilon^4\, u\derIIII(r_i)
+ O(\epsilon^6) \, ,
\eeq
which depends only on even derivatives of $u$.

The Noumerov function $z(r)$ is defined as
\beq
\nonumber
z(r_i)
\equiv u(r_i) - {\wf{1\over 12}}\epsilon^2\, u\derII(r_i) \, .
\eeq
By (\ref{eq:a4}) we can rewrite this as
\beq
z(r_i)
= \left( 1 - {\wf{1\over 12}}\epsilon^2\, W(r_i) \right) \, u(r_i).
\eeq
Note the $O(\epsilon^4)$ term drops out of the following Taylor
expansion
\beqa
z(r_{i+1})+ z(r_{i-1})
&=&
(u(r_{i+1})+ u(r_{i-1}))
-{\wf{1\over12}}\epsilon^2\, (u\derII(r_{i+1})+ u\derII(r_{i-1}))
\nonumber \\
&=&
2\, u(r_i)
+ \epsilon^2\, u\derII(r_i)
+ {\wf{1\over 12}} \epsilon^4\, u\derIIII(r_i)
+ O(\epsilon^6)
\nonumber \\
& \phantom{=} &
-{\wf{1\over 12}} \epsilon^2\, \left(
2 u\derII(r_i)
+ \epsilon^2 u\derIIII(r_i)
+ {\wf{1\over 12}} \epsilon^4\, u\derIIIIII(r_i)
+ O(\epsilon^6)
\right)
\nonumber \\
&=&
2\, \left( u(r_i) -{\wf {1\over 12}} \epsilon^2\, u\derII(r_i)\right)
+\epsilon^2\, u\derII(r_i)
+O(\epsilon^6)
\nonumber \\
&=&
2\, z(r_i)
+\epsilon^2\, u\derII(r_i)
+O(\epsilon^6)
\nonumber \\
&=&
2\, z(r_i)
+\epsilon^2\, W(r_i) \, u(r_i)
+O(\epsilon^6)
\eeqa
so that, including terms to $O(\epsilon^5)$,
\beq
\label{eq:zit}
z(r_{i+1}) =
2\, z(r_i) -z(r_{i-1}) +\epsilon^2\, W(r_i)\, u(r_i)
\eeq
and, finally,
\beq
\label{eq:uit}
u(r_{i+1})=
\left(1-{\wf{1\over 12}}\epsilon^2\,W(r_{i+1})\right)^{-1} \,
z(r_{i+1}) \, .
\eeq

Thus, the Noumerov prescription leads to a value for
$u(r_{i+1})$, given $z(r_i)$, $z(r_{i-1})$, and the $W(r_i)$,
which is accurate to $O(\epsilon^5)$, and which requires a
knowledge of the potential $V(r)$ only at the points $r_i$.

To start this numerical proceedure we simply choose
$u(0)$ and $u(\epsilon)$ according to the appropriate boundary
conditions and then iterate with step size $\epsilon$.
As $\psi({\bf r})\propto u(r)/r$ is the solution of the 3--D
\schrod equation, and finite at $r=0$, we require $u(0)=0$.
The choice of
$u(\epsilon)$
is arbitrary, it merely sets the overall magnitude of the
wavefunction.
We use $u(\epsilon)=\epsilon^{\ell+1}$, where $\ell$ is the
orbital angular momentum, to get the ``correct'' behaviour of
$u\derI(0)$.

\goodbreak

\section{Finding the Physical Solutions}
\label{apndx:asymptotic}

In the last Appendix we described the Noumerov technique for
solving the \schrod equation.
The physically allowed solutions must obey the additional
constraint that, as $r\to \infty$, we must have $u(r) \to 0$.
In this Appendix we describe a short--cut for finding these physical
solutions with two different potential functions $V(r)$.

Consider a quantum particle $R$ of mass $m$ bound in a three
dimensional central potential $V(r)$.
The radial equation that must be satisfied is
equation~(\ref{eq:schr1})
\cite{merzbacher}
\beq
\label{eq:brad}
-{\hbar^2 \over {2 m}}\, u_l\derII (r)
+\left[
 {\hbar^2 \over {2 m}} {\ell (\ell +1) \over r^2}
+ V(r) 
\right] \, u_l(r)
= E \, u_l(r) \, ,
\eeq
with the boundary conditions $u_\ell(0)=u_\ell(\infty)=0$.
Here we will only be concerned with the large $r$ behaviour of
the wavefunction, so the $1/r^2$ centripital barrier term
can be ignored.
We can, therefore, choose $\ell = 0$ and drop the subscript
$\ell$ on $u_\ell(r)$ without restricting our conclusions.

Consider first the simple, but important, case of a central
square well potential defined by
\beq
V(r) = \cases{
V_{int} & if $ r < r_{\rm c} $ \cr
V_{ext} & if $ r \geq r_{\rm c} $ \cr}
\eeq
where, for the moment, we restrict $E$ to satisfy
$V_{int} < E < V_{ext}$.

For $r<r_{\rm c}$, $u(r)=u_{int}(r)$ obeys
\beqa
u_{int}\derII (r)
&=&-{2 m \over \hbar^2 } \, (E - V_{int} ) \, u_{int}(r)
\nonumber \\
&=& - k^2_{int} \, u_{int}(r)
\eeqa
where the momentum
\beq
k_{int} \equiv \sqroot{{2\, m \over \hbar^2}(E-V_{int})}
= \sqroot{-W(r)}
\eeq
is, by construction, real.
This has the general solution
\beq
u_{int} (r) =
 D_{int} \sin (k_{int} r)
+G_{int} \cos (k_{int} r) \, .
\eeq
The reason for choosing the unusual nomenclemature, $D_{int}$
and $G_{int}$, for the unknown coefficients will become apparent
below.
The boundary condition $u_{int}(0)=0$ automatically leads to
$G_{int}=0$.

For $r>r_{\rm c}$, $u(r)=u_{ext}(r)$ obeys
\beqa
u_{ext}\derII (r)
&=&-{2 m \over \hbar^2 } \, (E - V_{ext} ) \, u_{ext}(r)
\nonumber \\
&=& + \kappa^2_{ext} \, u_{ext}(r)
\eeqa
where 
\beq
\kappa_{ext} \equiv \sqroot{-{2\, m \over \hbar^2}(E-V_{ext})}
\eeq
is, again by construction, real.
This has the general solution
\footnote{
We introduce the functions $d(r)$ and $g(r)$ for
future convenience.
We will continue to express the large $r$ wavefunctions of
bound systems as
linear combinations of some exponentially decaying
function $d(r)$ plus some exponentially growing function $g(r)$.
We will re--define the
functional form of $d(r)$ and $g(r)$ as we go from the
bound square well problem to the linear confining potential, but
they will play identical roles in all bound cases.
For unbound systems $d$ and $g$ will represent,
respectively, the sin and cos functions.}
\beqa
\label{eq:b_ext}
u_{ext} (r)
&=&
 D_{ext} e^{-\kappa_{ext} r}
+G_{ext} e^{+\kappa_{ext} r}
\nonumber \\
\noalign{\vskip 4pt}
&=&
 D_{ext} d (r)
+G_{ext} g (r)
\, .
\eeqa

Analytically, one solves this problem by setting
$u_{int}(r_{\rm c})=u_{ext}(r_{\rm c})$ and
$u_{int}^\prime(r_{\rm c})=u_{ext}^\prime(r_{\rm c})$
to find the unknown co--efficients
$D_{int}$, $D_{ext}$, and $G_{ext}$.
The physically allowed solutions are those for which $G_{ext}=0$.
It turns out that the values of the $D$ and $G$
depend on the energy $E$ through $k_{int}$ and $\kappa_{ext}$, so
the determination of the eigenvalues and eigenfunctions simply
becomes a matter of finding those $E$'s for which $G_{ext}=0$.

Starting with a trial energy of $E=(V_{int}+\Delta E)$,
where $\Delta E \ll (V_{ext}-V_{int})$,
and iterating to find $u(r)$ numerically as
discussed in Appendix~A, one finds that $G_{ext}$ is large and
has the same sign as $u(\epsilon)$.
As the trial energy is increased the value of $G_{ext}$ falls
and eventually changes sign.
Whenever $G_{ext}$ vanishes we have an eigenenergy and an
eigenfunction.
Finding the zeros of $G(E)$ is a simple matter of interpolation
between $G(E_-)$ and $G(E_+)$, where $E_-<E_{eigen}<E_+$,
and can easily be done to whatever accuracy the computer allows.
We would, however, like to find an efficient algorithm for finding
$G$ at each energy.
We address this now.

We determine $u(r)$ by stepping out from $r=0$, using the
numerical techniques of Appendix~A, until $r$ enters the
classically forbidden region.
(This region, which begins at $r_{\rm c}$, is the point
where the
kinetic energy either vanishes or becomes negative.)
We record the value of the wavefunction at two nearby points
$r_\b > r_\a > r_{\rm c}$:
\beqa
\label{eq:ba}
u_{ext}(r_\a) &=& D_{ext}\, d(r_\a) + G_{ext}\, g(r_\a)
\hskip 1 cm {\rm and}
\nonumber \\
u_{ext}(r_\b) &=& D_{ext}\, d(r_\b) + G_{ext}\, g(r_\b)
\eeqa
It is then trivial, with
$u(r)=u_{ext}(r)$, to solve for $G_{ext}$
\beq
\label{eq:bG}
G_{ext}={{ u(r_a) \, d(r_\b) - u(r_b) \, d(r_\a)
}\over{ g(r_\a) \, d(r_\b) - g(r_\b) \, d(r_\a) }} \, ,
\eeq
and we can stop the numerical iteration.
It is easy to write code to
increase the test energy by $\Delta E$ until
the sign of $G$ changes, then decrease the test energy by
some fraction of $\Delta E$ until the sign of $G$ changes
again, and so on, until one obtains the eigenenergy to the
desired level of accuracy.
In practice we must check to be sure that changing the values
of $r_\a$ and $r_\b$ leaves the eigenenergies fixed.

When the test energy $E$ exceedes $V_{ext}$ we no longer have an
eigenenergy situation, the quantum
particle becomes unbound, all energies are allowed, and the
external wavefunction takes the form
\beqa
\label{eq:b_free}
u_{ext}(r)
&=&
 D_{ext} \sin(k_{ext} r) +G_{ext} \cos(k_{ext} r)
\nonumber \\
&=&
 D_{ext}\, d(r) +G_{ext}\, g(r)
\eeqa
where
\beq
k_{ext} = \sqroot{{2\, m \over \hbar^2}(E-V_{ext})}
\eeq
is real.
Notice we have again re--defined the functional forms of $d(r)$
and $g(r)$ without changing the role they play in the analysis.
Finding $u(r_a)=u_{ext}(r_a)$ and $u(r_b)=u_{ext}(r_b)$
at two different points just
outside $r_{\rm c}$ allows us to find $G_{ext}$, via
(\ref{eq:bG}), and $D_{ext}$ from
\beq
D_{ext}={{ u(r_a) \, g(r_\b) - u(r_b) \, g(r_\a)
}\over{ d(r_\a) \, g(r_\b) - d(r_\b) \, g(r_\a) }}\, .
\eeq

We can now find the phase shift $\delta$, due to the
scattering of an unbound particle from the 3--D square well
potential, by noting that
\beqa
u_{ext}(r)
&=&
 D_{ext} \sin(k_{ext} r) +G_{ext} \cos(k_{ext} r)
\nonumber \\
&=&
  N \sin(k_{ext} r + \delta)
\nonumber \\
&=&
  N \,\left[\, \sin(\delta) \cos(k_{ext} r)
  +   \cos(\delta) \sin(k_{ext} r) \right]
\eeqa
(with $N$ a normalization factor) from which
\beq
\label{eq:delta1}
\delta = {\rm atan} \left({G_{ext}\over D_{ext}} \right) \, .
\eeq

{\it Thus we use the same algorithm for finding the wavefunction
of a bound or scattering state};
we must simply define the $d(r)$ and $g(r)$ by
(\ref{eq:b_ext}) or
(\ref{eq:b_free}) respectively!
For bound states we must tune the energy so that $G_{ext}=0$;
whereas for scattering states all energies are permitted and we
use (\ref{eq:delta1}) to find $\delta (E)$.

In practice this proceedure is found to work beautifully, and the
fortran code
\cite{fortran} developed to carry out this algorithm
runs quickly on a modest unix workstation.

To solve the more realistic problem of a pure
linear confining potential we use the WKB approximation
\cite{merzbacher}.
In the classically forbidden region $r>r_{\rm c}$
(with $r_{\rm c} = E/b$ and $u(r)=u_{ext}(r)$)
we approximate the solution to
\beq
\label{eq:b3}
-{\hbar^2 \over {2 m}} u\derII (r) + br\, u (r) = E u (r)
\eeq
by
\beqa
\label{eq:b4}
u (r) |_{\asym}
&\approx&
D_{ext}\, e^{-y(r)}/\sqroot{\kappa(r)}+
G_{ext}\, e^{+y(r)}/\sqroot{\kappa(r)} \nonumber \\
\noalign{\vskip 2 pt}
&=&
D_{ext}\, d(r) +
G_{ext}\, g(r)
\eeqa
where
\beq
\label{eq:bk}
\kappa = \sqroot{-{2\, m \over \hbar^2}(E-br)}
\eeq
and
\beqa
\label{eq:by}
y(r)
&=&
\int^r
\sqroot{-{2\, m \over \hbar^2}(E-br)}
\, dx \nonumber \\
\noalign{\vskip 4 pt}
&=&
{2\over 3b} \,
\sqroot{-{2\, m \over \hbar^2}(E-br)^3}
\eeqa

It appears as though these large $r$ wavefunctions have a different
dimension than the previous large $r$ wavefunctions (see, for
example,
equation~(\ref{eq:b_free})),
which would require that these expansion coefficients
$D_{ext}$ and $G_{ext}$ have different dimensionality that the
previous $D_{ext}$ and $G_{ext}$.
The discrepancy arises, however, because the sin and cos terms in
equation~(\ref{eq:b_free})
have implicit factors, with dimension $1/\sqrt{{\rm fm}}$
and magnitude unity,
that are required to normalize the otherwise divergent
sin and cos functions to one particle per unit length.

Notice also that we have again re--defined the functional form of
$d(r)$ and $g(r)$ between
equations~(\ref{eq:b_free}) and (\ref{eq:b4}) without
changing the role they play in our analysis.

We can again write two nearby values of the wavefunction, as
in equations~(\ref{eq:ba}),
and find $G_{ext}$ by equation~(\ref{eq:bG}).
Knowing $G_{ext}$ makes it easy to find
the eigenvalues and eigenfunctions of equation~(\ref{eq:b3}).

To solve this problem when the differential equation is given
by (\ref{eq:schr1}) and (\ref{eq:hqqbar}) we note that, in the
classically forbidden region, we can make the approximation
\beq
{\hbar^2\over2\,\mu_{\qq}}\, {\ell_\qq(\ell_\qq+1) \over \rqq^2}
\, + V_\Rqq \, (\rqq)
\approx m_{\sss q} + m_{\sss \bar q} + C + br
\eeq
which allows us to use the same scheme as in the pure linear case,
except that $r_{\rm c}$ is now determined by the condition
\beq
{\hbar^2\over2\,\mu_{\qq}}\, {\ell_\qq(\ell_\qq+1) \over r_{\rm c}^2}
+ V_\Rqq \, (r_{\rm c})
=E \, ,
\eeq
and
\beq
\kappa = \sqroot{-{2\, m \over \hbar^2}
(E-m_{\sss q} -m_{\sss \bar q} - C -br)}
\eeq
while
\beqa
y(r)
&=&
\int^r
\sqroot{-{2\, m \over \hbar^2}
(E-m_{\sss q} -m_{\sss \bar q} - C -br)}
\, dx \nonumber \\
\noalign{\vskip 4 pt}
&=&
{2\over 3b} \,
\sqroot{-{2\, m \over \hbar^2}
(E-m_{\sss q} -m_{\sss \bar q} - C -br)^3}
\eeqa

In practice these approximate schemes must be monitored
to ensure that the solutions do not depend on the numerical
parameters $\epsilon$, $r_\a$, and $r_\b$ introduced.
The model and the experimental results are also inexact, and
the numerical errors are easily  rendered insignificant by
comparison.

\goodbreak

\section{Solving the Two--Channel Model}
\label{apndx:2cse}

In this Appendix we discuss techniques for solving
equation~(\ref{eq:ccm}) for the specific case of
$AB \to R \to AB$.
Here $A$, $B$, and $R$ are ordinary $\qq$ resonances, but $A$
and $B$ are assumed to be stable against strong decay while $R$
obviously is not, it can have a lifetime of
$O(10^{-23})$ seconds.

Our first step is eliminating one of the variables $\rAB$
and $\rqq$.
We take the SU(3) limit of $\rqq = 2\, \rAB$.
This substitution is easily implemented by writing
equation~(\ref{eq:ccm})
in the form, with $r=r_\AB$,
\beq
\label{eq:c1}
\left[ \, \begin{array}{cc}
H_\AB (r) & V_\ABR (r) \\
V_\ABR (r) & H_\Rqq (2r)
\end{array} \, \right] \,
\left[ \, \begin{array}{c}
u_\AB (r) \\ u_\Rqq (2r)
\end{array} \, \right]
= E
\, \left[ \,
\begin{array}{c}
u_\AB (r) \\ u_\Rqq (2r)
\end{array}
\, \right]\, .
\eeq
We further simplify this equation by replacing the $\qq$
wavefunction
$u_\Rqq (2r)$ with
$u_\R (r) \equiv u_\Rqq (2r)$:
\beq
\label{eq:c2}
\left[ \, \begin{array}{cc}
H_\AB (r) & V_\ABR (r) \\
V_\ABR (r) & H_\Rqq (2r)
\end{array} \, \right] \!
\left[ \, \begin{array}{c}
u_\AB (r) \\ u_\R (r)
\end{array} \, \right]
= E
\, \left[ \, \begin{array}{c}
u_\AB (r) \\ u_\R (r)
\end{array} \, \right]\, .
\eeq

We can write the above equation in a more compact form if we
first define the Hamiltonian matrix
\beq
\HM (r) =
\left[ \, \begin{array}{cc}
H_\AB (r) & V_\ABR (r) \\
V_\ABR (r) & H_\Rqq (2r)
\end{array} \, \right]
\eeq
and the vector radial wavefunction
\beq
\u (r)
\equiv
\left[ \, \begin{array}{c}
u_1 (r) \\ u_2 (r)
\end{array} \, \right]
\equiv
\left[ \, \begin{array}{c}
 u_\AB (r) \\ u_\Rqq (2r)
\end{array} \, \right]
\equiv
\left[ \, \begin{array}{c}
 u_\AB (r) \\ u_\R (r)
\end{array} \, \right]
\eeq
Explicitly,
\beq
H_\Rqq (2r) =
-{\hbar^2\over2\,\mu_{\qq}}\,{1\over 2^2}\,{d^2 \over dr^2 }
+{\hbar^2\over2\,\mu_{\qq}}\,{1\over 2^2}\,
{\ell_\qq(\ell_\qq+1) \over r^2}
\, + m_q \, + \, m_{\bar q} \, + V_\Rqq \, (2r)
\eeq
and, from
equation~(\ref{eq:vqqbar}),
\beqa
\label{eq:c4}
V_\Rqq \, (2r)=
& - &
\! \!
\Biggl[
\, {3 \over 4}\, b\, 2r
\, -{\alpha_c \over 2r}
\, -\alpha_g \, {\sigma_g^3 \over \pi^{\sss 3/2} }
\, e^{-\sigma_g^2 (2r)^2}
\, + {3 \over 4}\, C
\Biggr]
\, \FF
\nonumber \\
\noalign{\vskip 4 pt}
& - &
{8\pi \over 3} {\alpha_h \over m_q m_{\bar q}}
\, {\sigma_h^3 \over \pi^{\sss 3/2} }
\, e^{-\sigma_h^2 (2r)^2}
\, \FF
\ {\bf S}_q \cdot {\bf S}_{\bar q}
\, .
\eeqa
With all of this
equation~(\ref{eq:c1}) becomes
\beq
\HM (r)\, \u (r) = E\, \u (r) \, .
\eeq

{}From
equations~(\ref{eq:hqqbar}),
(\ref{eq:H_AB1}), and
(\ref{eq:V_ABR}),
we see that each component of $\HM$ consists of a function of
$r$ plus, in the diagonal entries, a derivative operator.
We define a new matrix $\WM$, whose entries are proportional to
these functions of r, and whose two indicies refer, respectively,
to the final and initial channels, by
\beq
\WM (r) =
\left[ \, \begin{array}{cc}
W_{11} (r) & W_{12} (r) \\
W_{21} (r) & W_{22} (r)
\end{array} \, \right] \, ,
\eeq
with channel 1 being the $AB$ system and channel 2 the
$\qq$ state $R$.
Explicitly,
\beqa
\label{eq:cW}
W_{11}(r)
& = &
{2\,\mu_\AB \over \hbar^2} \left(
{\hbar^2 \over 2\, \mu_\AB}\, {\ell_\AB(\ell_\AB+1) \over r^2}
+V_{\ABAB}(r)+m_\A+m_\B-E \right) \nonumber \\
\noalign{\vskip 4 pt}
W_{12}(r)
& = & {2\, \mu_\AB \over \hbar^2}\, V_\ABR(r) \nonumber \\
\noalign{\vskip 4 pt}
W_{22}(r)
& = & {2\,\mu_\qq \over \hbar^2} \ 2^2\,
\left({\hbar^2 \over 2\, \mu_\qq}\, {1\over2^2}\,
{\ell_\qq(\ell_\qq+1) \over r^2}
+V_{\qq}(2r)+m_\q+m_\qbar-E \right) \nonumber \\
\noalign{\vskip 4 pt}
W_{21}(r)
& = & {2\, \mu_\qq \over \hbar^2} \ 2^2 \,
V_{\ABR}(r)
 =  {\mu_\qq \over \mu_\AB} \ 2^2 \, W_{12}(r) \, ,
\eeqa
The factors of $2^2$ in (\ref{eq:cW}) effect the change from
$\rqq$ to $2\,r$.

We can now write equation~(\ref{eq:c1}) as
\beq
\label{eq:CCM}
\u^{\prime \prime}(r) = \, \WM (r)\, \u(r) \, .
\eeq
(\ref{eq:CCM}) is equivalent in every way to
equation~(\ref{eq:ccm}), but it
lends itself to a matrix formuation of the Noumerov technique and
is easily adapted to generate numerical solutions.

Analagous to Appendix \ref{apndx:noumerov}, we define the vector
Noumerov function as
\beq
\z (r) = \u (r)
- {\wf{1\over 12}}\epsilon^2\, \u\derII (r)\\
\eeq
so, by
(\ref{eq:CCM}),
\beq
\label{eq:defz}
\z (r)
=\left(
\IM-{\wf{1\over 12}}\epsilon^2\,\WM (r) \, \right) \u (r)
\eeq
where $\IM$ is the unit matrix.

When we make these functions depend on the discrete variable
$r_i$ and Taylor expand $\u(r_i)$ we find, in analogy with
equation~(\ref{eq:aupm}),
\beq
\u (r_{i+1})+ \u (r_{i-1})
=2\, \u (r_i)
+ \epsilon^2\, \u\derII(r_i)
+ {\wf{1\over12}}\epsilon^4\, \u\derIIII(r_i)
+ O(\epsilon^6)
\eeq
and, in analogy with equation~(\ref{eq:zit}),
\beq
\label{eq:czit}
\z (r_{i+1}) =
2\, \z (r_i)
-\z (r_{i-1})
+\epsilon^2\, \WM (r_i)\, \u (r_i)
+O(\epsilon^6)\, .
\eeq
To find $\u (r_{i+1})$, which is the wavefunction we are
interested in, we
simply invert equation~(\ref{eq:defz}):
\beq
\label{eq:findu}
\u(r_{i})
= \left(
\IM-{\wf{1\over 12}}\epsilon^2\,\WM (r_{i}) \right)^{-1} \,
\z(r_{i})
\eeq
Therefore, knowing
$\u (r_i)$,
$\z (r_i)$,
$\z (r_{i-1})$,
and the
$\WM (r_i)$,
allows us to calulate
$\u (r_{i+1})$:
{\it i.e.}, we have an algorithm for finding $\u (r)$, given
$\u (0)$,
$\u (\epsilon)$, and
$\WM (r)$, which is exact to
$O(\epsilon^5)$.

Having found the mathematical solutions to equation~(\ref{eq:c1}) we
must now, as in Appendix \ref{apndx:asymptotic},
find the physically allowed solutions, namely those
which have $u_\qq(\rqq)$ vanish as $\rqq\to \infty$.

As in the case of the single channel \schrod equation,
the iterated numerical solution will have, generally, both
exponentially growing and decaying components in the classically
forbidden region.
In the single channel case we were able to adjust the test
energy of the bound state to find the eigenenergy and the
physical wavefunctions.
In the two--channel problem, however, the energy is set by the
total energy which $A$ and $B$ bring to the interaction, and has
any value greater than $m_\A + m_\B$.
We must, therefore, find a new criterion for defining the physical
wavefunctions.

The two--channel large $r$ wavefunction is
\beq
\label{eq:lr2}
\u(r)\vert_{\asym} =
\left[ \, \begin{array}{c}
D_{1}\sin(k_1r)+ G_{1}\cos(k_1r) \\
D_{2}\, {e^{-y_2(r)} / \sqroot{\kappa_2(r)}} +
G_{2}\, {e^{+y_2(r)} / \sqroot{\kappa_2(r)}}
\end{array} \, \right]
\eeq
with
\beqa
k_1 &=&
\sqrt{{2\,\mu_\AB \over \hbar^2}(E-m_\A-m_\B)},
\nonumber \\ \noalign{\vskip 4 pt}
\kappa_2(r)
&=&
\sqroot{-{2\, \mu_2 \over \hbar^2}
(E-m_{\sss q} -m_{\sss \bar q} -C - br)}
\nonumber \\ \noalign{\vskip 4 pt}
{\rm and} \hskip 2 true cm
y_2(r)
&=&
{2\over 3b} \,
\sqroot{-{2\, \mu_2 \over \hbar^2}
(E-m_{\sss q} -m_{\sss \bar q} -C - br)^3}
\eeqa
and $r_{\rm c}$ defined as the point at which the $\qq$
kinetic energy vanishes, corresponding to the boundary between
the classically allowed and forbidden regions.
The upper component describes two free mesons with kinetic energy
\hbox{$(E-m_\A-m_\B)$} and the lower component, written in
analogy with equation~(\ref{eq:b4}), describes a
$\qq$ state bound by a linear confining potential.
The unknown coefficients $D_{1}$, $D_{2}$, $G_{1}$, and
$G_{2}$ are determined using the techniques discussed in
Appendix~B.

It is obvious that the solution we want has $G_{2}=0$ in 
(\ref{eq:lr2}).
Suppose we generate two different solutions of
equation~(\ref{eq:CCM}) according to two different sets of initial
conditions:
\beqa
\u_1(0)
&=&
\left[ \, \begin{array}{c} 0 \\ 0 \end{array} \, \right]
\hskip 0.8 true cm
{\rm and}
\hskip 0.8 true cm
\u_1(\epsilon)=
\left[ \, \begin{array}{c}
\epsilon^{(\ell_1+1)} \\ 0 \end{array} \, \right]
\, ,
\\
\noalign{\noindent{\rm and}} \nonumber \\
\u_2(0)
&=&
\left[ \, \begin{array}{c} 0 \\ 0 \end{array} \, \right]
\hskip 0.8 true cm
{\rm and}
\hskip 0.8 true cm
\u_2(\epsilon)=
\left[ \, \begin{array}{c} 0 \\
\epsilon^{(\ell_2+1)} \end{array} \, \right]
\, ,
\eeqa
where $\ell_j$ is the orbital angular momentum of the $j^{th}$
channel.
These will generate the aymptotic solutions
\beq
\label{eq:c19}
\u_1(r)\, \vert_{\asym} =
\left[ \, \begin{array}{c}
D_{11}\sin(k_1r)+ G_{11}\cos(k_1r) \\
D_{21}\, {e^{-y_2(r)} / \sqroot{\kappa_2(r)}} +
G_{21}\, {e^{+y_2(r)} / \sqroot{\kappa_2(r)}}
\end{array} \, \right]
\eeq
and
\beq
\label{eq:c20}
\u_2(r)\, \vert_{\asym} =
\left[ \, \begin{array}{c}
D_{12}\sin(k_1r)+ G_{12}\cos(k_1r) \\
D_{22}\, {e^{-y_2(r)} / \sqroot{\kappa_2(r)}} +
G_{22}\, {e^{+y_2(r)} / \sqroot{\kappa_2(r)}}
\end{array} \, \right] \, ,
\eeq
where we have added an index to the $D$ and $G$ relative to
equation~(\ref{eq:lr2}) to accomodate our having two
solutions.
By defining a matrix of wavefunctions as
\beq
\UM (r) \equiv [\, \u_1(r) \ \ \u_2(r) \, ] \, ,
\eeq
(\ref{eq:c19}) and (\ref{eq:c20})
can be written
\beqa
\label{eq:cum}
\UM (r) \, |_{\asym}
&=&
\left[ \, \begin{array}{cc}
\sin (k_1 r) & 0 \\
0 & {e^{-y_2(r)} / \sqroot{\kappa_2(r)}}
\end{array} \, \right] \!
\left[ \, \begin{array}{cc}
D_{11} & D_{12} \\
D_{21} & D_{22}
\end{array} \, \right] \nonumber \\
\noalign{\vskip 2 pt}
&+&
\left[ \, \begin{array}{cc}
\cos (k_1 r) & 0 \\
0 & {e^{+y_2(r)} / \sqroot{\kappa_2(r)}}
\end{array} \, \right] \!
\left[ \, \begin{array}{cc}
G_{11} & G_{12} \\
G_{21} & G_{22}
\end{array} \, \right] \, .
\eeqa

For future convenience we define the matricies
\beq
\DM \ \equiv
\left[ \, \begin{array}{cc}
D_{11} & D_{12} \\
D_{21} & D_{22}
\end{array} \, \right] , \
\GM \ \equiv
\left[ \, \begin{array}{cc}
G_{11} & G_{12} \\
G_{21} & G_{22}
\end{array} \, \right] , \ {\rm and} \
\CM \ \equiv \ \GM^{-1} \, =
\left[ \, \begin{array}{cc}
C_{11} & C_{12} \\
C_{21} & C_{22}
\end{array} \, \right] \, ,
\eeq
where we know that $\CM$ exists because
$\u_1(r)$ and $\u_2(r)$ are linearly independent solutions.

By noting that
\beq
(\HM (r) - \IM E) \UM \, =0
\eeq
we see that $\UM \CM$ is also a solution of the two--channel \schrod
equation:
\beq
(\HM (r) - \IM E) \UM \CM \, =0 \, .
\eeq

Using equation~(\ref{eq:cum}) we can write
\beqa
\label{eq:cuphy}
\UM (r)\CM \, |_{\asym}
&=&
\left[ \, \begin{array}{cc}
\sin (k_1 r) & 0 \\
0 & {e^{-y_2(r)} / \sqroot{\kappa_2(r)}}
\end{array} \, \right] \!
\left[ \, \begin{array}{cc}
D_{11} & D_{12} \\
D_{21} & D_{22}
\end{array} \, \right] \!
\left[ \, \begin{array}{cc}
C_{11} & C_{12} \\
C_{21} & C_{22}
\end{array} \, \right] \nonumber \\
\noalign{\vskip 2 pt}
&+&
\left[ \, \begin{array}{cc}
\cos (k_1 r) & 0 \\
0 & {e^{+y_2(r)} / \sqroot{\kappa_2(r)}}
\end{array} \, \right] \, .
\eeqa
We note that the first column of
(\ref{eq:cuphy})
is exactly the solution that we are seeking, it has
no exponentially growing component!

It is useful to simplify this equation by introducing yet another
matrix:
\beq
\FM\, =
\left[ \, \begin{array}{cc}
F_{11} & F_{12} \\
F_{21} & F_{22}
\end{array} \, \right]
=
\left[ \, \begin{array}{cc}
D_{11} & D_{12} \\
D_{21} & D_{22}
\end{array} \, \right] \!
\left[ \, \begin{array}{cc}
C_{11} & C_{12} \\
C_{21} & C_{22}
\end{array} \, \right]
=\DM\,\CM
\eeq
where
\beq
F_{ij}=\sum_{k=1}^2 D_{ik}\, C_{kj}\, .
\eeq
The physical solution $\u_\P(r)$ is
the first column of the solution matrix $\UM\! (r)\! \CM$:
\beq
\label{eq:cphys}
\u_\P (r) |_\asym
=
[ \, \UM (r)|_{\asym} \CM \, ]_1 \,
=
N\,
\left[ \, \begin{array}{c}
F_{11} \sin (k_1 r) + \cos (k_1 r) \\
F_{21} {e^{-y_2(r)} / \sqroot{\kappa_2(r)}}
\end{array} \, \right] \, ,
\eeq
where $N$ is a normalization factor.
This solves equation~(\ref{eq:ccm})!

If we normalize the $AB$ wavefunction to unit amplitude (per
unit length) then
\beq
\label{eq:uphys}
\u_\P (r) |_\asym
=
\left[ \, \begin{array}{c}
\sin (k_1 r + \delta^{2ch})
\\
N_\R \,
{e^{-y_2(r)} / \sqroot{\kappa_2(r)}}
\end{array} \, \right] \, ,
\eeq
where
\beq
N_\R =
{F_{21} \over \sqroot{F_{11}^2 + 1}} \, ,
\eeq
and the 2--channel phase shift induced as $AB$ forms $R$ and $R$
then decays to $AB$ is
\beq
\label{eq:delta2}
\delta^{2ch} = {\rm atan} \left({1\over F_{11}}
\right) \, .
\eeq

By solving for $\delta^{2ch}(E)$ at a range of energies beginning at
$AB$ threshold we can compare the predictions of the coupled
channel equations directly with the canonical Breit--Wigner
phase shift for $AB \to R \to AB$.
This comparison is made in
FIG.~6. and discussed in
Section~\ref{sec:twochannel}.

\def\pk{\hskip 0.3 true cm}

Since
\beq
\UM (0) \, =
\left[ \, \begin{array}{cc}
\pk 0 \pk & \pk 0 \pk \\
\pk 0 \pk & \pk 0 \pk
\end{array} \, \right]
\hskip 1 true cm
{\rm and}
\hskip 1 true cm
\UM (\epsilon) \, =
\left[ \, \begin{array}{cc}
\epsilon^{(\ell_1+1)} & 0 \\
0 & \epsilon^{(\ell_1+1)}
\end{array} \, \right]
\eeq
and
\beq
\u_\P (r)
=
[ \, \UM (r) \CM \, ]_1 \, ,
\eeq
we can generate these physical solutions numerically by
starting with the intitial conditions
\beq
\label{eq:pic0}
\u_\P (0)
=
[ \, \UM (0) \CM \, ]_1
=
\left[ \, \begin{array}{cc}
\pk 0 \pk & \pk 0 \pk \\
\pk 0 \pk & \pk 0 \pk
\end{array} \, \right]_1
=
\left[ \, \begin{array}{c}
0  \\
0 
\end{array} \, \right]
\eeq
and
\beq
\label{eq:pic1}
\u_\P (\epsilon)
=
[ \, \UM (\epsilon) \CM \, ]_1
=
\left[
\left[ \, \begin{array}{cc}
\epsilon^{(\ell_1+1)} & 0 \\
0 & \epsilon^{(\ell_1+1)}
\end{array} \, \right]
\left[ \, \begin{array}{cc}
C_{11} & C_{12} \\
C_{21} & C_{22}
\end{array} \, \right]
\right]_1
=
\left[ \, \begin{array}{c}
C_{11}\, \epsilon^{(\ell_1 +1)}  \\
C_{21}\, \epsilon^{(\ell_2 +1)} 
\end{array} \, \right] \, .
\eeq

We can numerically integrate the lower component squared of
equation~(\ref{eq:uphys}) to obtain the probability of finding the
$AB\lra R$ system in $R$.
A plot of this probability as a function of the scattering
energy $E$ is shown in
FIG.~9.
Note that the probability of $R$ being present in the central
region never vanishes,
even when the phase shift of $AB$--in to $AB$--out equals an
integer multiple of $\pi$.
This is because $AB$--in is a shell of inward--moving $AB$
probability amplitudes rather than two columnated beams.

\goodbreak


\section{Solving the Multichannel Model}
\label{apndx:mcse}

\def\bfa{{\bf a}}
\def\bfA{{\bf A}}
\def\bfe{{\bf \hat e}}
\def\rma{{\rm a}}
\def\rmA{{\rm A}}
\def\rmC{{\rm C}}
\def\rmS{{\rm S}}
\def\FMtwo{\FM^{{\sss (2)}}}
\def\IMtwo{\IM^{{\sss (2)}}}
\def\rootvAB{\sqrt{4 \pi v_1}}
\def\rootvCD{\sqrt{4 \pi v_2}}

At this point we turn to a more general multichannel problem
that contains essentially all the special cases we must
consider;
the two--channel equation (C1) is generalized to the
four--channel equation for $AB \lra CD \lra Q \lra R$:
\beq
\label{eq:c7}
\left[ \, \begin{array}{cccc}
H_\AB (r)  & V_\ABCD (r) & V_\ABQ (r) & V_\ABR (r) \\
V_\ABCD (r) & H_\CD (r) & V_\CDQ (r) & V_\CDR (r) \\
V_\ABQ (r) & V_\CDQ (r) & H_\Qqq (2r) & V_\QR (r) \\
V_\ABR (r) & V_\CDR (r) & V_\QR (r) & H_\Rqq (2r)
\end{array} \, \right] \!
\left[ \, \begin{array}{c}
u_\AB (r) \\ u_\CD (r) \\ u_\Q (r) \\ u_\R (r)
\end{array} \, \right]
= E \!
\left[ \, \begin{array}{c}
u_\AB (r) \\ u_\CD (r) \\ u_\Q (r) \\ u_\R (r)
\end{array} \, \right]\, .
\eeq
Here $AB$ and $CD$ represent two
different meson--meson final states, with
$m_\A + m_\B \leq m_\C + m_\D$,\footnote{
As an example, to study the mass splitting
caused by different decay channel couplings,
and the $D/S$ amplitude ratio,
for the reaction
$(\omega\pi)_{\sss s} \lra
 (\omega\pi)_{\sss d} \lra
 b_1(1235)$
we would construct a three--channel equation with
$m_\A + m_\B = m_\C + m_\D$.}
and $Q$ and $R$ represent two meson resonances with $m_\Q < m_\R$.

In (D1) those elements already introduced are
$H_\AB$ and $H_\CD$, given by
equation~(\ref{eq:H_AB1}), or its analogue,
$H_\Qqq$ and $H_\Rqq$, given by
equations~(\ref{eq:schr1}) and
(\ref{eq:vqqbar}), or their analogues, and
$V_\ABQ$, $V_\ABR$, $V_\CDQ$, and $V_\CDR$, given by
equation~(\ref{eq:V_ABR}), or its analogues.
There are two new elements,
$V_\ABCD$ is the quark exchange potential
reviewed in Section \ref{sec:qx},
and $V_\QR$, represents possible direct meson--meson mixing
such as $\eta-\eta^\prime$ or $\omega-\phi$ mixing,
which we set = 0 here.
Allowing this entry to be non--zero is one way to mock--up the
presence of a glueball state.
\footnote{A more realistic way would be to form a separate
glueball channel analogous to a $q\bar q$ channel.
In our non--relativistic formulation this would require modelling
the diagonal glueball Hamiltonian in terms of ``constituent"
valance gluons
\cite{barnes81}.}

Analogous to the two--channel case (Appendix~C), we define a
4x1 vector wavefunction
\beq
\u (r)
\equiv
\left[ \, \begin{array}{c}
u_1 (r) \\ u_2 (r) \\ u_3 (r) \\ u_4 (r)
\end{array} \, \right]
\equiv
\left[ \, \begin{array}{c}
 u_\AB (r) \\ u_\CD (r) \\ u_\Q (r) \\ u_\R (r)
\end{array} \, \right]
=
\left[ \, \begin{array}{c}
u_\AB (r) \\ u_\CD (r) \\ u_\Qqq (2r) \\ u_\Rqq (2r)
\end{array} \, \right]
\eeq
and a 4x4 matrix $\WM(r)$ with entries given by the direct
analogues of equations~(\ref{eq:cW}).

\def\px{\hskip 0.061 true cm}
\def\pz{\hskip 0.428 true cm}
\def\pq{\hskip 0.000 true cm}

We again form a matrix of, in this case, 4 solutions $\u_j(r)$,
with $j=1,4$, and use the Numerov technique to find these
wavefunction given 4 orthogonal initial conditions:
\beqa
\UM (0)
&=&
\left[ \, \begin{array}{cccc}
\px\u_1(0)\px &\px\u_2(0)\px &\px\u_3(0)\px &\px\u_4(0)\px \\
\end{array} \, \right] \nonumber \\
\noalign{\vskip 2 pt}
&=&
\left[ \, \begin{array}{cccc}
\pz 0 \pz & \pz 0 \pz & \pz 0 \pz & \pz 0 \pz \\
\pz 0 \pz & \pz 0 \pz & \pz 0 \pz & \pz 0 \pz \\
\pz 0 \pz & \pz 0 \pz & \pz 0 \pz & \pz 0 \pz \\
\pz 0 \pz & \pz 0 \pz & \pz 0 \pz & \pz 0 \pz
\end{array} \, \right] \nonumber \\
\noalign{\noindent{\rm and}} \nonumber \\
\UM (\epsilon)
&=&
\left[ \, \begin{array}{cccc}
\pq \epsilon^{(\ell_1+1)} \pq & \pq 0 \pq & \pq 0 \pq & \pq 0 \pq \\
\pq 0 \pq & \pq \epsilon^{(\ell_2+1)} \pq & \pq 0 \pq & \pq 0 \pq \\
\pq 0 \pq & \pq 0 \pq & \pq \epsilon^{(\ell_3+1)} \pq & \pq 0 \pq \\
\pq 0 \pq & \pq 0 \pq & \pq 0 \pq & \pq \epsilon^{(\ell_4+1)} \pq
\end{array} \, \right] \, ,
\eeqa
where each column forms a separate and linearly independent
solution of the four dimensional form of
equation~(\ref{eq:CCM}) and $\ell_j$ is the orbital angular momenta of
the $j^{th}$ channel.

Each of these solutions will have one of two possible large $r$
expansions for the $CD$ system, depending on whether the
energy $E$ is smaller than or larger than $m_\C + m_\D$,
so that, respectively, $CD$ is either bound or free.

If $E<m_\C+m_\D$ then $CD$ may exist as a bound state confined
to the
central region of the interaction, but not as a free state.
(Recall that we are thinking of $A$, $B$, $C$, and $D$ as $\qq$
states which are stable against strong decays.)
In the large $r$ regions all $CD$ potentials have vanished and
the $CD$ confinement is due solely to conservation of energy.
Therefore, when we have found the wavefunctions by
iteration from $r=0$ to $r_{\sss b} > r_{\sss a} > r_{\rm c}$
using the techniques described in
Appendix~\ref{apndx:2cse},
we must represent the large $r$ bound $CD$
wavefunction as the sum of an exponentially decaying part and an
exponentially growing part, in direct analogy with the square
well discussion in
Appendix~\ref{apndx:asymptotic}.
The $AB$ system remains free and
the resonances $Q$ and $R$ are bound by linear confining
potentials as in
Appendix~\ref{apndx:2cse}.
For the $j^{th}$ solution these large $r$ wavefunctions are
\beq
\u_j(r)\, \vert_{\asym} =
\left[ \, \begin{array}{c}
D_{1j}\sin(k_1r)+ G_{1j}\cos(k_1r) \\
D_{2j}\, e^{-\kappa_2r}+ G_{2j}\, e^{+\kappa_2r} \\
D_{3j}\, {e^{-y_3(r)} / \sqroot{\kappa_3(r)}} +
G_{3j}\, {e^{+y_3(r)} / \sqroot{\kappa_3(r)}} \\
D_{4j}\, {e^{-y_4(r)} / \sqroot{\kappa_4(r)}} +
G_{4j}\, {e^{+y_4(r)} / \sqroot{\kappa_4(r)}}
\end{array} \, \right]
\eeq
with real `momenta'
$\kappa_2 = \sqroot{(2\, \mu_\CD / \hbar^2)(E-m_\C-m_\D)}$.
$\kappa_2$ plays the same role that $\kappa_{ext}$ played in
Appendix~\ref{apndx:asymptotic}.
All other quantities are as defined as in
Appendix~\ref{apndx:2cse}, and
the $D$ and $G$ coefficients are found as described in
Appendix~\ref{apndx:asymptotic}.

With these four solutions we go through exactly the same
proceedure as discussed in
Appendix~\ref{apndx:2cse},
but now all matricies are 4x4 and $\FM$ has its 16 components
defined by
\beq
\label{eq:defF4}
[\,\FM\ ]_{ij}=F_{ij}=\sum_{k=1}^4D_{ik}\,C_{kj}\,.
\eeq
As in Appendix~\ref{apndx:2cse}, we find the large $r$
form of the physical solution $\u_P$ to be
the first column of the solution matrix $\UM\! (r) \CM$:
\beqa
\label{eq:duphy}
\u_\P (r) |_\asym =
[ \, \UM (r)|_{\asym} \CM \, ]_1 \, =
N \,
\left[ \, \begin{array}{c}
F_{11} \sin (k_1 r) + \cos (k_1 r) \\
F_{21} e^{-\kappa_2r} \\
F_{31} \, e^{-y_3(r)} / \sqroot{\kappa_3(r)} \\
F_{41} \, e^{-y_4(r)} / \sqroot{\kappa_4(r)}
\end{array} \, \right]
\eeqa
where $N$ is a normalization factor.

The phase shift of the $AB$ state resulting from
$AB \to ( AB\lra CD\lra Q\lra R) \to AB$
scattering process is
\beq
\delta^{4ch} = {\rm atan} \left({1\over F_{11}} \right) \, .
\eeq

Since we are below the energy for inelastic scattering to
$CD$, all the $AB$ amplitude flowing into the interaction region
must also flow out as $AB$.
We choose, for later convenience, to normalize the
spherical inward--moving external wavefunction to unit flux
crossing an imaginary sphere surrounding the interaction region,
which leads to.
\def\onein{{\sss 1;{\rm in}}}
\beq
\sqrt{4\pi v_\AB}\, (u_\ABin^\ast u_\ABin) \, =\,
\sqrt{4\pi v_1}\, (u_\onein^\ast u_\onein) \, =\, 1\, ,
\eeq
where $v_\AB =v_1= \sqrt{2\,(E-m_\A-m_\B)/\mu_\AB}$ is the
``reduced" velocity of $AB$ (channel 1) in.
This has the effect of implementing conservation of particle
number
\cite{blatt}.
With this choice
\beq
\u_\P (r) |_\asym
=
{1\over\rootvAB}\,
\left[ \, \begin{array}{c}
\, \sin (k_1 r + \delta^{2ch}) \\
N_2 \, e^{-\kappa_2r} \\
N_3 \, {e^{-y_3(r)} / \sqroot{\kappa_3(r)}} \\
N_4 \, {e^{-y_4(r)} / \sqroot{\kappa_4(r)}} \\
\end{array} \, \right]
\eeq
where, we see from (\ref{eq:duphy}), that, for $j=2,3,4,$
\beq
N_j = {F_{j1}\over\sqroot{F_{11}^2 + 1}} \, .
\eeq

When $E>m_\C+m_\D$ we are in a very different regime,
$CD$ is now a free state and we can describe inelastic
scattering processes such as $AB \to CD$.
We use the same proceedure as above to find 4 solutions to (D1),
only now we must represent the large $r$
wavefunction of the $j^{th}$ solution by
\beq
\u_j(r)\, \vert_{\asym} =
\left[ \, \begin{array}{c}
D_{1j}\, \sin(k_1r)+ G_{1j}\, \cos(k_1r) \\
D_{2j}\, \sin(k_2r)+ G_{2j}\, \cos(k_2r) \\
D_{3j}\, {e^{-y_3(r)} / \sqroot{\kappa_3(r)}} +
G_{3j}\, {e^{+y_3(r)} / \sqroot{\kappa_3(r)}} \\
D_{4j}\, {e^{-y_4(r)} / \sqroot{\kappa_4(r)}} +
G_{4j}\, {e^{+y_4(r)} / \sqroot{\kappa_4(r)}}
\end{array} \, \right]
\eeq
where the momentum
$k_2 = \sqroot{(2\, \mu_\CD / \hbar^2)(E-m_\C-m_\D)}$ is real.
In this case there are now two physically allowed solutions,
corresponding to the first two columns of the matrix
$\UM\!(r)\CM$.
We find, in direct analogy with
(\ref{eq:cphys}), that these two solutions are
\beq
\label{eq:d12}
\u_{{\sss P1}} (r) |_\asym =
[ \, \UM (r)|_{\asym} \CM \, ]_1 \, =
\left[ \, \begin{array}{c}
F_{11}\,  \sin (k_1 r) + \cos (k_1 r) \\
F_{21}\,  \sin(k_2r) \\
F_{31}\,  e^{-y_3(r)} / \sqroot{\kappa_3(r)} \\
F_{41}\,  e^{-y_4(r)} / \sqroot{\kappa_4(r)}
\end{array} \, \right]
\eeq
and
\beq
\label{eq:d13}
\u_{{\sss P2}} (r) |_\asym =
[ \, \UM (r)|_{\asym} \CM \, ]_2 \, =
\left[ \, \begin{array}{c}
F_{12}\,  \sin (k_1 r) \\
F_{22}\,  \sin(k_2r)+ \cos (k_2 r) \\
F_{32}\,  e^{-y_3(r)} / \sqroot{\kappa_3(r)} \\
F_{42}\,  e^{-y_4(r)} / \sqroot{\kappa_4(r)}
\end{array} \, \right]
\eeq

Of course, now every linear combination of
(\ref{eq:d12}) and (\ref{eq:d13}) is also
a solution of the multichannel equation.
We are interested in solving for two specific linear
combinations, one corresponding to an incident state of pure
$AB$, which models the inelastic process $AB \to AB$ and $CD$, and the
other with pure $CD$ in, for $CD \to AB$ and $CD$.

For pure $AB$ incident we wish to find the complex coefficients
$\rma_{11}$ and $\rma_{21}$ so that
\goodbreak
\beqa
\label{eq:ABin}
\u_\ABin (r) |_\asym
&=&
\rma_{11} \,
\u_{{\sss P1}} (r) |_\asym \, + \,
\rma_{21} \,
\u_{{\sss P2}} (r) |_\asym \nonumber \\
\noalign{\vskip 2 pt}
&=&
\left[ \, \begin{array}{c}
(F_{11}\,\rma_{11}+F_{12}\,\rma_{21})\,\sin(k_1\,r) +
 \rma_{11}\,\cos(k_1\,r) \\
(F_{21}\,\rma_{11}+F_{22}\,\rma_{21})\,\sin(k_2\,r) +
 \rma_{21}\,\cos(k_2\,r) \\
(F_{31}\,\rma_{11}+F_{32}\,\rma_{21})\,
e^{-y_3(r)}/\sqroot{\kappa_3(r)} \\
(F_{41}\,\rma_{11}+F_{42}\,\rma_{21})\,
e^{-y_4(r)}/\sqroot{\kappa_4(r)}
\end{array} \, \right] \nonumber \\
\noalign{\vskip 2 pt}
&=&
\hskip 0.82 true cm
\left[ \, \begin{array}{c}
1\ e^{-ik_1 r} + \rmA_{11} \ e^{+ik_1 r} \\
0\ e^{-ik_2 r} + \rmA_{21} \ e^{+ik_2 r} \\
(F_{31}\,\rma_{11}+F_{32}\,\rma_{21})\,
e^{-y_3(r)}/\sqroot{\kappa_3(r)} \\
(F_{41}\,\rma_{11}+F_{42}\,\rma_{21})\,
e^{-y_4(r)}/\sqroot{\kappa_4(r)}
\end{array} \, \right]
\eeqa
where $\rmA_{11}$ and $\rmA_{21}$ are the complex coefficients
of the outward--moving $AB$ and $CD$ systems, respectively, and are
to be determined from the numerical solution.
All effects of the intermediate resonances $Q$ and $R$ on the
scattering states will be realized in the coefficients
$\rmA_{11}$ and $\rmA_{21}$ of the outward--moving waves.
We shall address the issue of normalization when we relate these
solutions to the $S$--matrix below.

We see that only the upper two components of
(\ref{eq:ABin}) will help us solve for the 4 complex unknowns
$\rma_{11}$, $\rma_{21}$,
$\rmA_{11}$, and $\rmA_{21}$.
As $e^{\pm i k r} = \cos(kr) \pm i \sin(kr)$, equating the
coefficients of $\cos(k_1r)$, $\sin(k_1r)$, $\cos(k_2r)$, and
$\sin(k_2r)$ leads to the four complex equations
\beqa
\label{eq:d4c}
F_{11}\,\rma_{11}+F_{12}\,\rma_{21}
&=&-i+i\rmA_{11}\ ,\nonumber\\
\noalign{\vskip 4 pt}
F_{21}\,\rma_{11}+F_{22}\,\rma_{21}
&=&i\rmA_{21}\ ,\nonumber\\
\noalign{\vskip 4 pt}
\rma_{11}&=&1+\rmA_{11}\ ,\hskip 1 true cm {\rm and}\nonumber\\
\noalign{\vskip 4 pt}
\rma_{21}&=&\rmA_{21}\, .
\eeqa
By defining
\beq
\FMtwo =
\left[ \, \begin{array}{cc}
F_{11} & F_{12} \\
F_{21} & F_{22}
\end{array} \, \right]
, \hskip 0.3 true cm
\bfa_1 =
\left[ \, \begin{array}{c}
\rma_{11} \\
\rma_{21}
\end{array} \, \right]
, \hskip 0.3 true cm
\bfA_1 =
\left[\,\begin{array}{c}\rmA_{11}\\ \rmA_{21}\end{array}\,\right]
, \hskip 0.3 true cm
{\rm and}
\hskip 0.3 true cm
\bfe_1 =
\left[ \, \begin{array}{c} 1 \\ 0 \end{array} \, \right]
\eeq
equations (\ref{eq:d4c}) can be written in matrix form
\beqa
\FMtwo \, \bfa_1 &=& -i \bfe_1 + i \bfA_1
\nonumber\\ \noalign{\vskip 2 pt}
\bfa_1 &=& \bfe_1 + \bfA_1 \, .
\eeqa
(Note that, although $\FMtwo$ is now a 2x2 matrix, it still
depends on $\u_3(r)$ and $\u_4(r)$ because the sum in
equation~(\ref{eq:defF4}) runs from 1 to 4.)
{}From this one can show that the required complex co--efficients are
\beq
\label{eq:d_sola}
\bfa_1
=2
\left[\IMtwo -i\,\FMtwo\,\right]
\left[\,\IMtwo+\FMtwo\,\FMtwo\,\right]^{-1}
\bfe_1
\eeq
and the resulting complex amplitudes of the outward--moving
waves are
\beq
\label{eq:d_solA}
\bfA_1
=2
\left[\IMtwo -i\,\FMtwo\,\right]
\left[\,\IMtwo+\FMtwo\,\FMtwo\,\right]^{-1}
\bfe_1
-\bfe_1
\eeq
where $\IMtwo$ is the 2x2 unit matrix.
These completely specify specify the physical wavefunctions for
$AB$ incident.


Solving this problem for pure $CD$ incident simply requires that
we find the complex coefficients
$\rma_{12}$ and $\rma_{22}$ so that
\beqa
\label{eq:CDin}
\u_\CDin (r) |_\asym
&=&
\rma_{12} \,
\u_{{\sss P1}} (r) |_\asym \, + \,
\rma_{22} \,
\u_{{\sss P2}} (r) |_\asym \nonumber \\
\noalign{\vskip 2 pt}
&=&
\left[ \, \begin{array}{c}
(F_{11}\,\rma_{12}+F_{12}\,\rma_{22})\,\sin(k_1\,r) +
 \rma_{12}\,\cos(k_1\,r) \\
(F_{21}\,\rma_{12}+F_{22}\,\rma_{22})\,\sin(k_2\,r) +
 \rma_{22}\,\cos(k_2\,r) \\
(F_{31}\,\rma_{12}+F_{32}\,\rma_{22})\,
e^{-y_3(r)}/\sqroot{\kappa_3(r)} \\
(F_{41}\,\rma_{12}+F_{42}\,\rma_{22})\,
e^{-y_4(r)}/\sqroot{\kappa_4(r)}
\end{array} \, \right]
\nonumber \\
\noalign{\vskip 2 pt}
&=&
\hskip 0.82 true cm
\left[ \, \begin{array}{c}
0\ e^{-ik_1 r} + \rmA_{12} \ e^{+ik_1 r} \\
1\ e^{-ik_2 r} + \rmA_{22} \ e^{+ik_2 r} \\
(F_{31}\,\rma_{12}+F_{32}\,\rma_{22})\,
e^{-y_3(r)}/\sqroot{\kappa_3(r)} \\
(F_{41}\,\rma_{12}+F_{42}\,\rma_{22})\,
e^{-y_4(r)}/\sqroot{\kappa_4(r)}
\end{array} \, \right]
\eeqa
where $\rmA_{12}$ and $\rmA_{22}$ are the complex coefficients
of the outward--moving $AB$ and $CD$ systems, respectively.

The solution to this problem is found by defining
\beq
\bfa_2 =
\left[ \, \begin{array}{c}
\rma_{12} \\
\rma_{22}
\end{array} \, \right]
,\,
\bfA_2 =
\left[\,\begin{array}{c}\rmA_{12}\\ \rmA_{22}\end{array}\,\right]
, \hskip 0.3 true cm
{\rm and}
\hskip 0.3 true cm
\bfe_2 =
\left[ \, \begin{array}{c} 0 \\ 1 \end{array} \, \right]
\eeq
and then by replacing $\bfe_1$ in
equations~(\ref{eq:d_sola}) and (\ref{eq:d_solA})
by $\bfe_2$, which leads to solutions for $\bfa_2$ and $\bfA_2$:
\beq
\bfa_2
={2\over\rootvCD}
\left[\IMtwo -i\,\FMtwo\,\right]
\left[\,\IMtwo+\FMtwo\,\FMtwo\,\right]^{-1}
\bfe_2
\eeq
and
\beq
\bfA_2
=2
\left[\IMtwo -i\,\FMtwo\,\right]
\left[\,\IMtwo+\FMtwo\,\FMtwo\,\right]^{-1}
\bfe_2
-\bfe_2
\eeq

Since we now know all four wavefunctions for the four possible
two--channel scattering processes $AB \lra CD$ we can find the
two--channel $S$--matrix from the parameterization
\beq
\label{eq:Smatrix}
\SM=
\left[ \, \begin{array}{cc}
\rmS_{11} & \rmS_{12} \\
\rmS_{21} & \rmS_{22}
\end{array} \, \right] \, = \,
\left[ \, \begin{array}{cc}
\eta \, e^{2\, i\, \delta_1} &
i\, [1-\eta^2]^{{\sss 1/2}} \, e^{i (\delta_1 + \delta_2)} \\
i\, [1-\eta^2]^{{\sss 1/2}} \, e^{i (\delta_1 + \delta_2)} &
\eta \, e^{2\, i\, \delta_2}
\end{array} \, \right] \, .
\eeq

To express the parameters $\eta$, $\delta_1$, and
$\delta_2$ in terms of the coefficients of the outward--moving
wavefunctions we need to consider wavefunction normalizations
and conservation of particle flux.

Notice, from equation~(\ref{eq:ABin}), that the external part of the
wavefunction consists of a shell of inward--moving $AB$ and a
shell of outward--moving $AB$ and $CD$.
We can express this as
\beq
\left[ \, \begin{array}{c}
\{1\}\,e^{-ik_1 r} \\
\{0\}\,e^{-ik_2 r}
\end{array} \, \right]
\longrightarrow
\left[ \, \begin{array}{c}
\{\rmA_{11}\}\,e^{+ik_1 r}\\
\{\rmA_{21}\}\,e^{+ik_2 r}
\end{array} \, \right] \, .
\eeq
Conservation of particle flux requires that the number of
inward--moving and outward--moving
particles must be equal in any given unit of time.
We can realize this by normalizing the inward--moving wave to unit
flux, \ie multiplying through by a factor of ${1/\rootvAB}$,
which leads
to
\beqa
\label{eq:ABscat}
\left[ \, \begin{array}{c}
\{1\}\,e^{-ik_1 r}\\
\{0\}\,e^{-ik_2 r}
\end{array} \, \right]
{1\over\rootvAB}
&\longrightarrow&
\left[ \, \begin{array}{c}
\{\rmA_{11}\}\,e^{+ik_1 r}\\
\{\rmA_{21}\}\,e^{+ik_2 r}
\end{array} \, \right]
{1\over\rootvAB}
\nonumber \\
\noalign{\vskip 2 pt}
&=&
\left[ \, \begin{array}{c}
\{\rmA_{11}\}\,e^{+ik_1 r}/\rootvAB\\
\{\rmA_{21}\sqrt{{v_2/v_1}}\}\,e^{+ik_2 r}/\rootvCD
\end{array} \, \right] \, .
\eeqa

Thus the multichannel quark model takes an inward--moving unit
flux normalized $AB$ system having an amplitude of 1 into a
linear combination of outward--moving unit normalized $AB$ and
$CD$ with complex amplitudes $\rmA_{11}$ and
$\rmA_{21}\sqrt{{v_2/v_1}}$ respectively.
(As $CD$ is heavier than $AB$ (by construction), they have lower
relative velocity
at a given total energy, and so the wavefunction describing
their relative separation must have greater density per unit
volume to satisfy conservation of particle flux.)

For pure inward--moving $CD$ (\ref{eq:CDin}) says
\beq
\label{eq:CDscat}
\left[ \, \begin{array}{c}
\{0\}\,e^{-ik_1 r}\\
\{1\}\,e^{-ik_2 r}
\end{array} \, \right]
{1\over\rootvCD}
\longrightarrow
\left[ \, \begin{array}{c}
\{\rmA_{12}\sqrt{{v_1/v_2}}\}\,e^{+ik_1 r}/\rootvAB\\
\{\rmA_{22}\}
\,e^{+ik_2 r}/\rootvCD
\end{array} \, \right] \, .
\eeq

{}From (\ref{eq:Smatrix}), (\ref{eq:ABscat}) and (\ref{eq:CDscat}) we
find that
\beq
\SM=
\left[ \, \begin{array}{cc}
\rmS_{11} & \rmS_{12} \\
\rmS_{21} & \rmS_{22}
\end{array} \, \right] \, = \,
\left[ \, \begin{array}{cc}
\rmA_{11} & \rmA_{12}\sqrt{{v_1/v_2}} \\
\rmA_{21}\sqrt{{v_2/v_1}} & \rmA_{22}
\end{array} \, \right]
\eeq
so that, with (\ref{eq:Smatrix})
\beqa
\delta_1
&=&
{1\over2}{\rm atan}
\left({{\rm Im\{A_{11}\}\over Re\{A_{11}\}}}\right) \, ,
\nonumber \\
\noalign{\vskip 4 pt}
\delta_2
&=&
-{\rm atan} \left({{\rm Re\{A_{21}\}\over Im\{A_{21}\}}}\right)
-\delta_1 \, , \hskip 1 true cm {\rm and}
\nonumber \\
\noalign{\vskip 4 pt}
\eta
&=&
\sqrt{{\rm (Im\{A_{11}\})^2 + (Re\{A_{11}\})^2}} \, ,
\eeqa
which solves (D1) for two free two--meson states.

We can also write the cross section, in millibarnes, for an
initial state $i$ to
scatter into a final state $f$ as
\cite{blatt}
\beq
\sigma_{fi} = { 20 \pi \over k_f^2 } \left[
({\rm Re\{S}_{fi}\}-\delta_{fi})^2 +
{\rm (Im\{S}_{fi}\})^2 \right]
\eeq
where $k_f$ is the momentum of the outward--moving state in inverse
fermi and $\delta_{fi}$ is 1 when $f=i$ and zero otherwise.

The generalization of this proceedure to arbitrary numbers of
external states $A_1B_1$, $A_2B_2$, $A_3B_3$, ...
and intermediate resonance states is a straightforward
exercise left to the reader.
We will, however, only require only one or two resonance states
until we get to the glueball, hybrid, or
charmed quark sectors: in isospin 0 we have the
$\sqrt{1\over2}(u\bar u + d\bar d)$ and $s\bar s$ states,
while in I=1 and $1/2$ we have only one possible $\qq$ resonance.
To date the largest system solved is the seven channel, two
resonance,
$\pi\pi,\ K\bar K,\ \eta\eta,\ \eta\eta^\prime,\ \eta^\prime
\eta^\prime,\ f_0,\ f_0^\prime$ system describing $s$--wave
$\pi\pi$ scattering
\cite{jw_cebaf}.

\vfill\eject



\begin{figure}[htbp]
\begin{center}
\leavevmode
\hbox{%
\epsfxsize=4.0 in
\epsffile{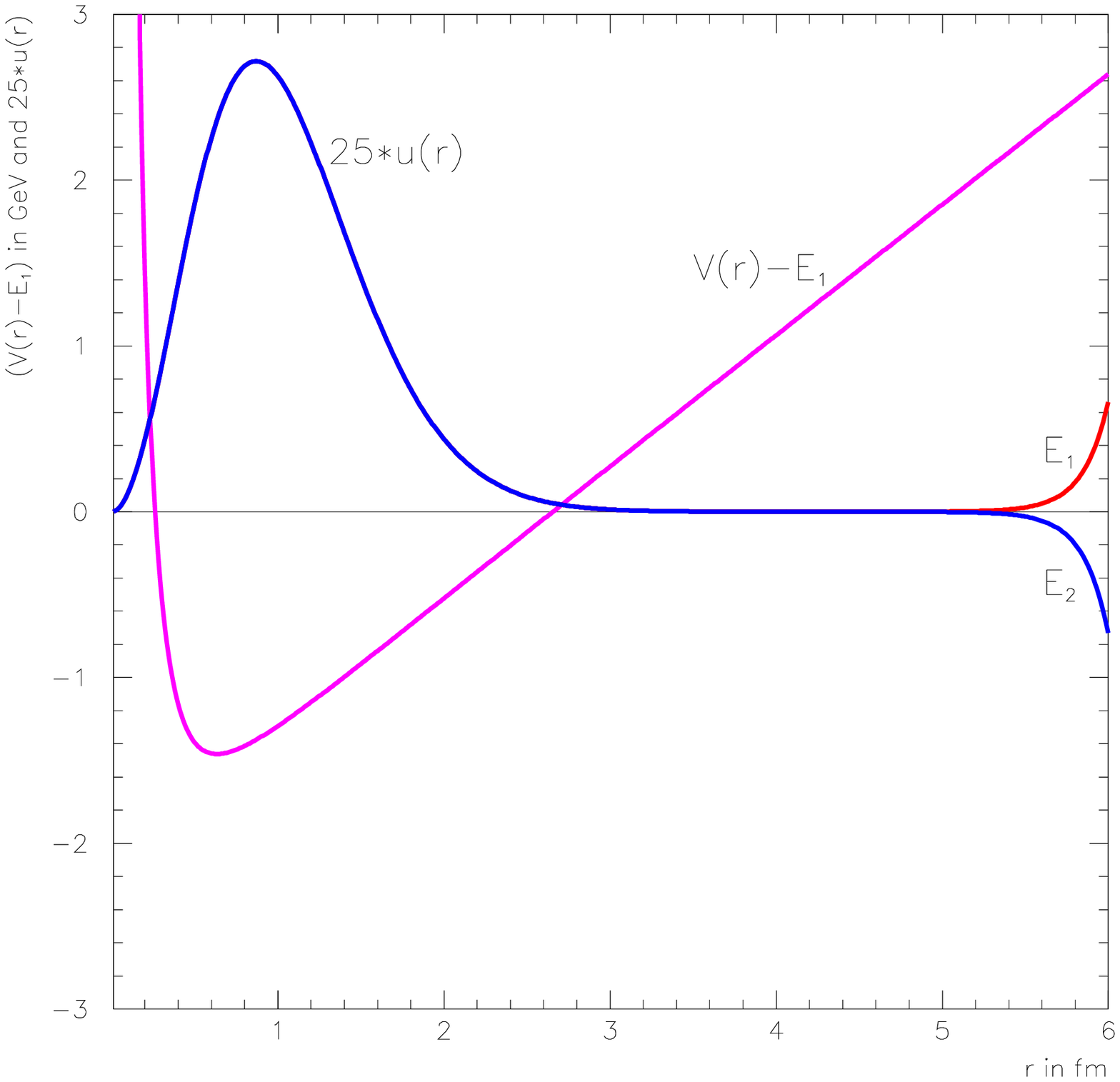}
}\end{center}
\caption { Here the $V_\qq$ potential has the quantum
numbers of the $K_J$,
$\ell_{\sss u\bar s}=1$, and
$s_{\sss u\bar s}=1$.
$(V_\qq(r)-E_1)$ and $25\, u_\qq(r)$
are plotted for two solutions of
equation~(5), with trial energies
$E_1=1.4299235012$~GeV and
$E_2=1.4299235014$~GeV.
The actual eigenenergy lies between these values.
We plot $(V_\qq(r)-E_1)$ rather than $V_\qq(r)$
so that the zero of $u(r)$ and the zero of available kinetic energy
both lie on the $r$ axis.
We pick the wavefunction normalization of 25 simply
so we can see the wavefunction and the
potential together.
}
\label{fig:int_wfn1}
\end{figure}

\begin{figure}[htb]
\centerline {\epsfxsize=4.0 true in
\epsffile{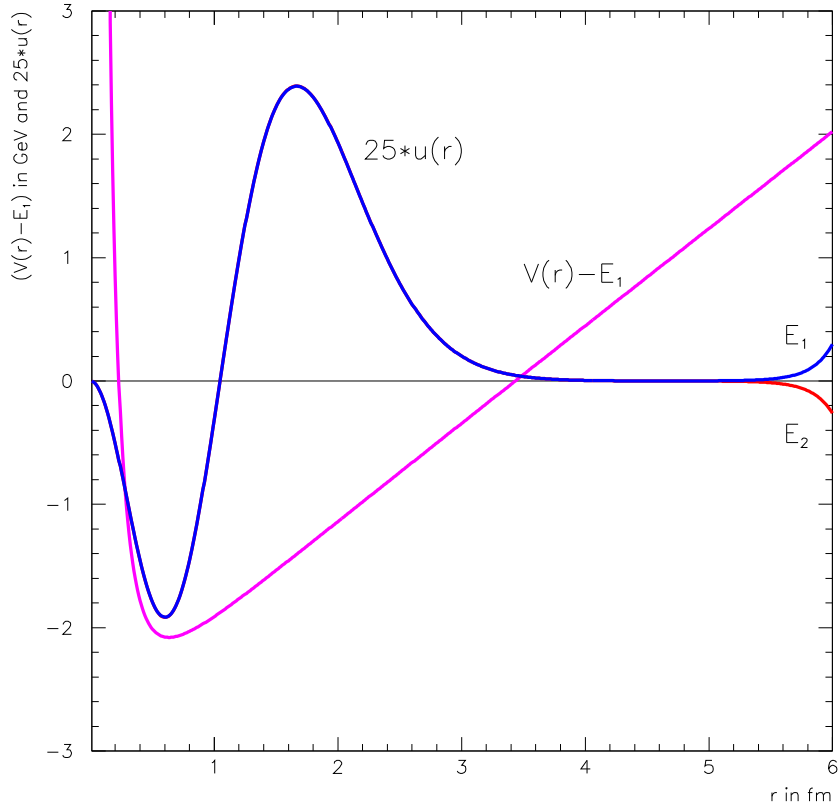} }
\caption { The same as
FIG.~1., but for the first radially excited $K_J$ state.
The trial energies are
$E_1=2.047471763$~GeV and
$E_2=2.047471768$~GeV.
}
\label{fig:int_wfn2}
\end{figure}

\begin{figure}[htb]
\label{fig:qld1}
\centerline {\epsfxsize=4.0 true in
\epsffile{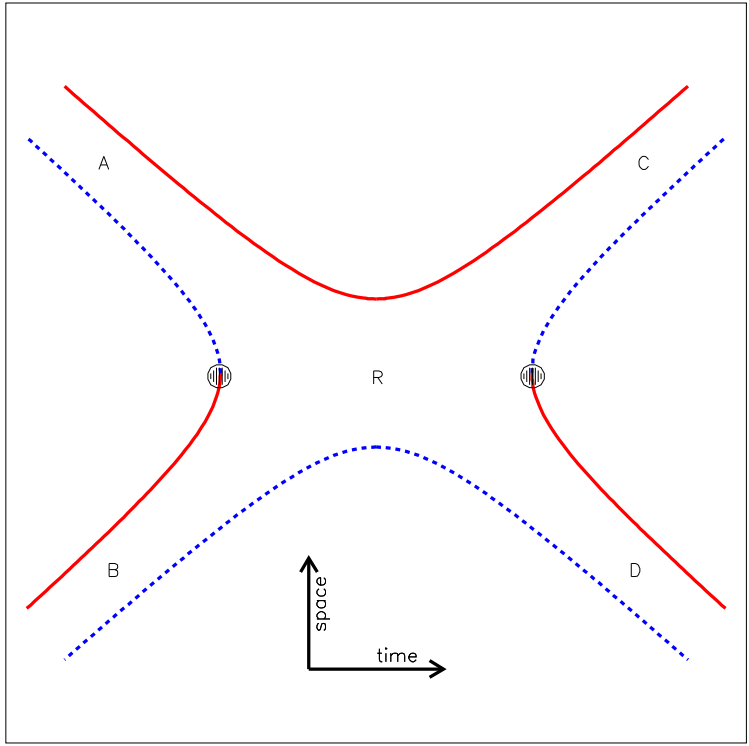} }
\caption { A quark line diagram for $s$--channel meson resonance
formation in meson--meson scattering.
The solid lines are quarks, the dashed lines are anti-quarks,
and the shaded dots are $\qq$ annihilation or creation
vertecies.
Understanding the properties of the resonance $R$ in terms of
the underlying $\qq$ interactions and the $AB \to R \to CD$
couplings is a goal of modern
particle physics.
As the lifetime of $R$ is typically $O(10^{-23})$ seconds it is
clear that the properties of $R$ may depend significantly on
its couplings to $AB$ and $CD$.
}
\end{figure}

\begin{figure}[htb]
\label{fig:qld2}
\centerline {\epsfxsize=4.0 true in
\epsffile{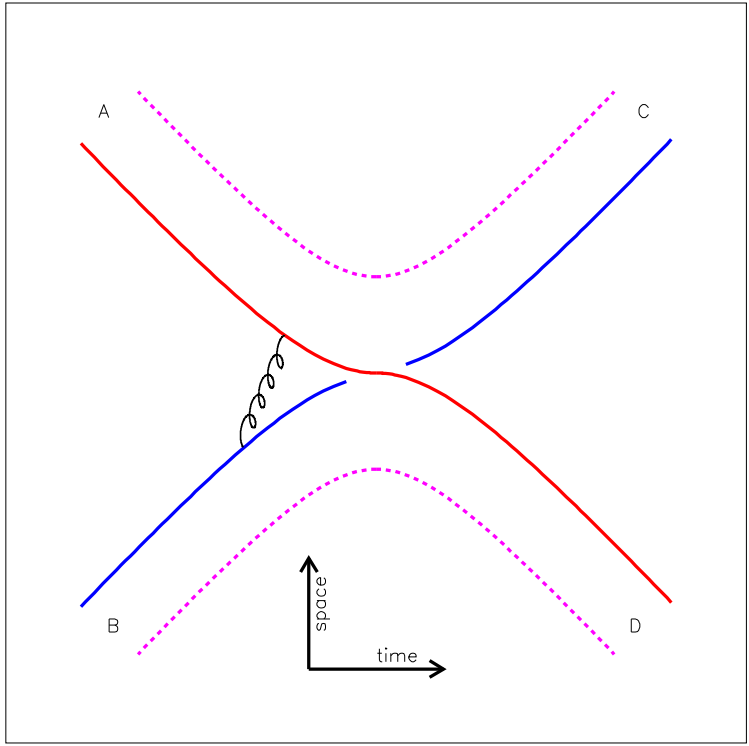} }
\caption { A quark line diagram for one--gluon exchange (the
curly line) followed by quark exchange
in meson--meson scattering.
This diagram leads to non--resonant
hadronic range van der Waals type intermeson potentials which
have gaussian shapes and ranges of about 0.5 fm.}
\end{figure}

\begin{figure}[htb]
\label{fig:rel_sep}
\centerline {\epsfxsize=4.0 true in
\epsffile{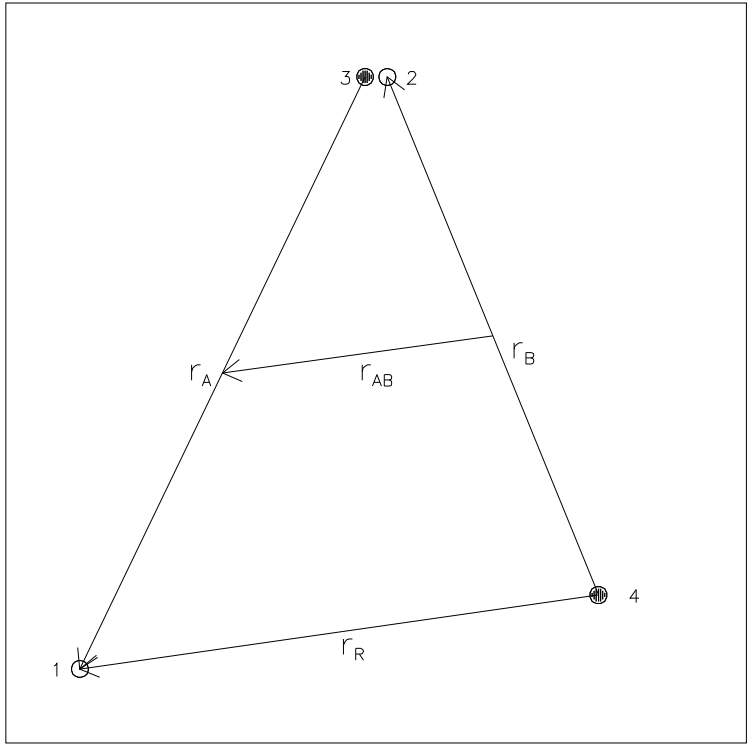} }
\caption {
The relation between $\rAB$ and $r_\R=\rqq$.
Here
$A=q_1\bar q_3$,
$B=q_2\bar q_4$, and
$R=q_1\bar q_4$.
Clearly $r_\R = 2\, \rAB$ when all quarks are of equal mass.
}
\end{figure}

\begin{figure}[htb]
\label{fig:bw_ccp}
\centerline {\epsfxsize=4.0 true in
\epsffile{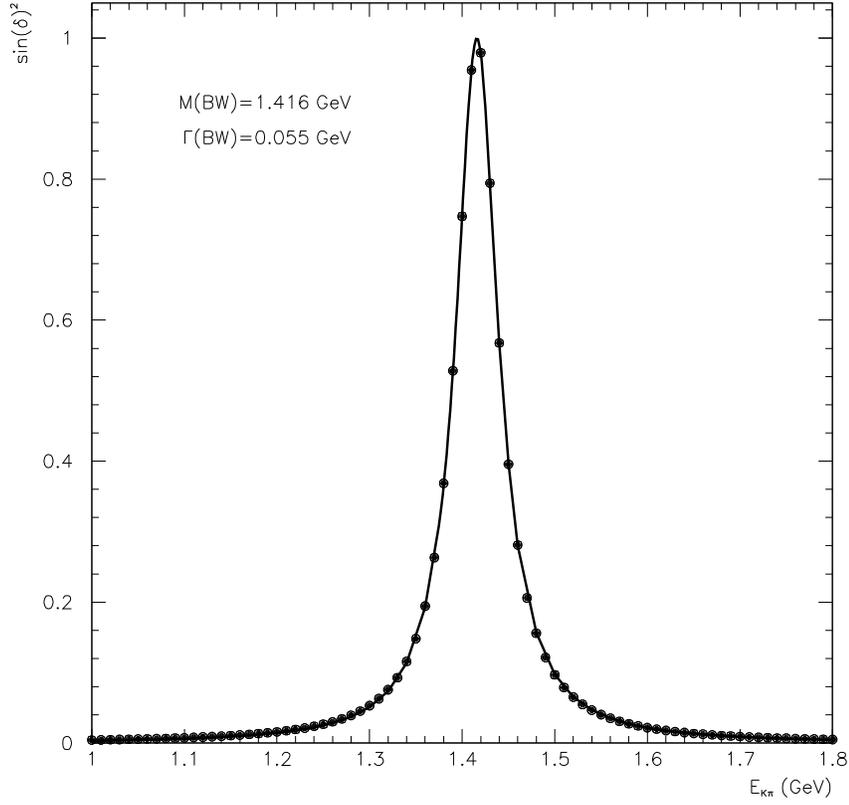} }
\caption
{
Compare
$\sin^2 (\delta^{\BW}_{{\sss K\pi}})$ (the dots) and
$\sin^2 (\delta^{\sss 2ch}_{{\sss K\pi}})$ (the solid line)
for $s$--wave $K\pi$ scattering.
Here
$\delta^{\BW}_{{\sss K\pi}}$ is the
Breit--Wigner phase shift and
$\delta^{\sss 2ch}_{{\sss K\pi}}$ (the solid line)
is the phase shift found from the two--channel quark model,
with the quark exchange potential in the quark model
equal set to zero.
The agreement between the two descriptions is excellent.
}
\end{figure}

\begin{figure}[htb]
\label{fig:delta_g}
\centerline {\epsfxsize=4.0 true in
\epsffile{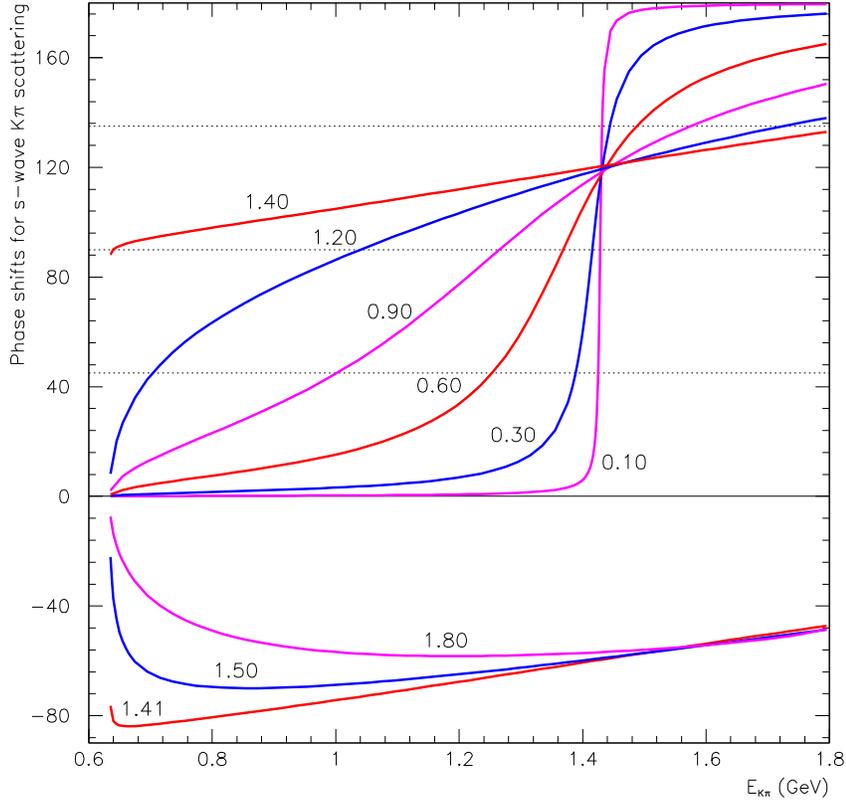} }
\caption
{The variation in the 2--channel $K\pi$ $s$--wave scattering
phase shift $\delta^{2ch}$ as $g$ varies.
The horizontal dotted lines mark $45^\circ$, $90^\circ$, and
$135^\circ$, and the numbers closest to the phase shifts give
the value of $g$ in GeV.
Note the departure from the Breit--Wigner phase shift description
as the width increases, and the onset of a $K\pi$ bound
state as $g$ grows larger than 1.40~GeV.
We define the width as the energy difference between the energy at
$\delta^{2ch}=135^\circ$ and
$\delta^{2ch}=45^\circ$,
and the resonance energy as the energy at which 
$\delta^{2ch}=90^\circ$.
}
\end{figure}

\begin{figure}[htb]
\label{fig:delta_kpi}
\centerline {\epsfxsize=4.0 true in
\epsffile{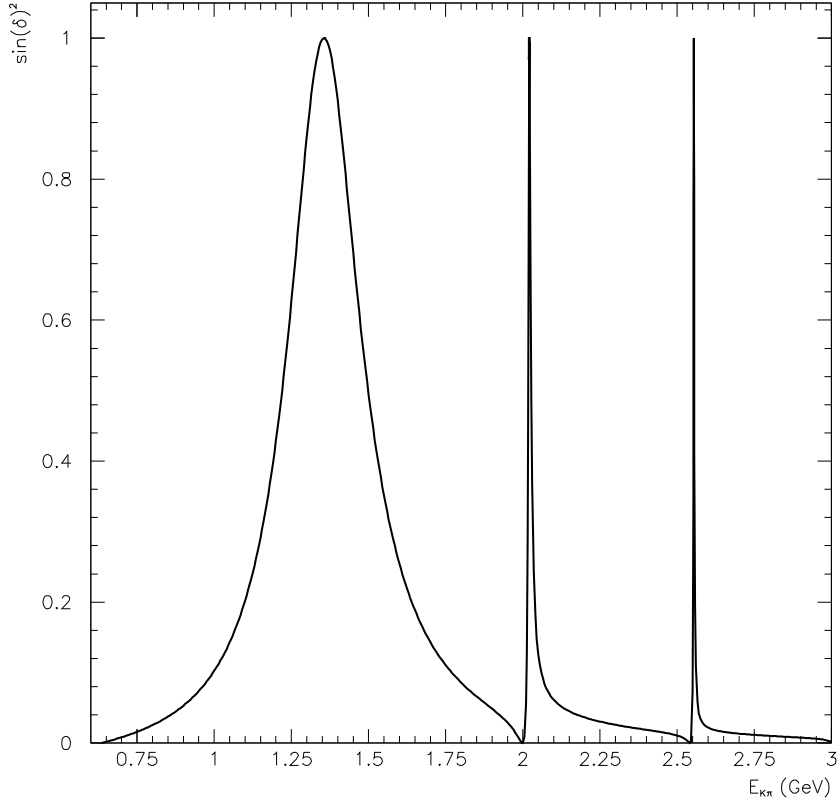} }
\caption{
$\sin^2 (\delta^{\sss 2ch}_{{\sss K\pi}})$, from
$s$--wave $K\pi$ scattering, is plotted from threshold to
$E_{K\pi}=3.0$~GeV, with
$g=0.60$~GeV and $\Gamma^{2ch}_{K_0}=280$~MeV.
The peaks at 1.356, 2.020, and 2.553~GeV correspond to the ground
state, the first radial excitation, and the second radial
excitation of
the $L=1$, $S=1$, $J=0$ strangeness $\pm 1$ resonance.
The narrowing of the excited states depends on the form of the
$\qq$ annihilation/creation verticies and requires further study.
Note that heavier states have more open channels, and that here we
are looking only at the partial width into a single channel.
}
\end{figure}

\begin{figure}[htb]
\label{fig:mag_bnd}
\centerline {\epsfxsize=4.0 true in
\epsffile{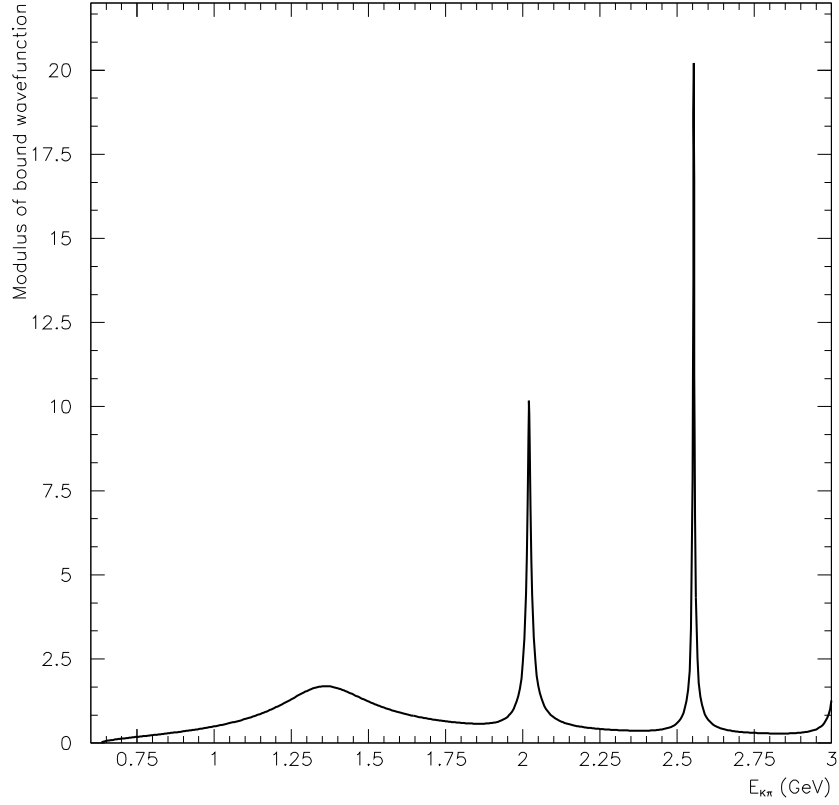} }
\caption
{The probability,
$\int_0^\infty | u_{K_{\sss 0}}^\ast u_{K_{\sss 0}} |^2\, dr$,
that the $s$-wave $K\pi$ scattering state is
in a central $\qq$ resonance with $K_{\sss 0}$ quantum numbers,
plotted as a
function of the $K\pi$ scattering energy.
Note that some $u\bar s$
is always present in the central region.
The location and widths of these peaks are directly related to
the mass and width of the $\qq$ resonances.
}
\end{figure}

\begin{figure}[htb]
\label{fig:kj}
\centerline {\epsfxsize=4.0 true in
\epsffile{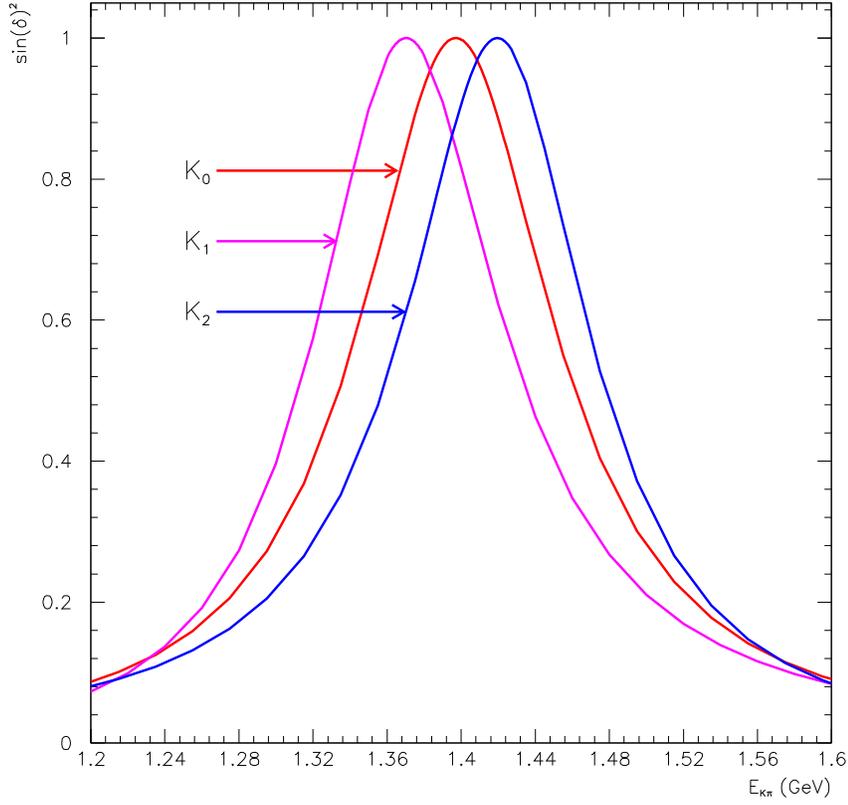} }
\caption
{Resonance masses of the strangeness $\pm 1$ $\ell_\qq=1$,
$s_\qq=1$, $J=0,1,2$ triplet as $J$ varies.
The $K_{0}$ and $K_{2}$ are obtained from the $s$-- and
$d$--wave scattering of a $K\pi$ system respectively, while the
$K_{1}$ is obtained from $s$--wave scattering of a $K^\ast
\pi$ pair.
The resonances masses are 1.397, 1.370, and 1.419~GeV
for $J=0$, 1, and 2 respectively.
The $K_2 - K_0$ mass splitting of 22~MeV arises from the
different vertex couplings to $K\pi$ whereas the $K_0 - K_1$ mass
difference of 17~MeV arises from $s$--wave couplings to
different final states.
The underlying $K_0,\ K_1,\ K_2$ masses, in this version of the
\naive quark model are degenerate since we have
omitted spin--orbit couplings in the Hamiltonian.
The strength of the vertex couplings have been adjusted so each
resonance has the same width in order to reduce the number of
differences and clarify the discussion.
}
\end{figure}

\begin{figure}[htb]
\label{fig:bw_m}
\centerline {\epsfxsize=4.0 true in
\epsffile{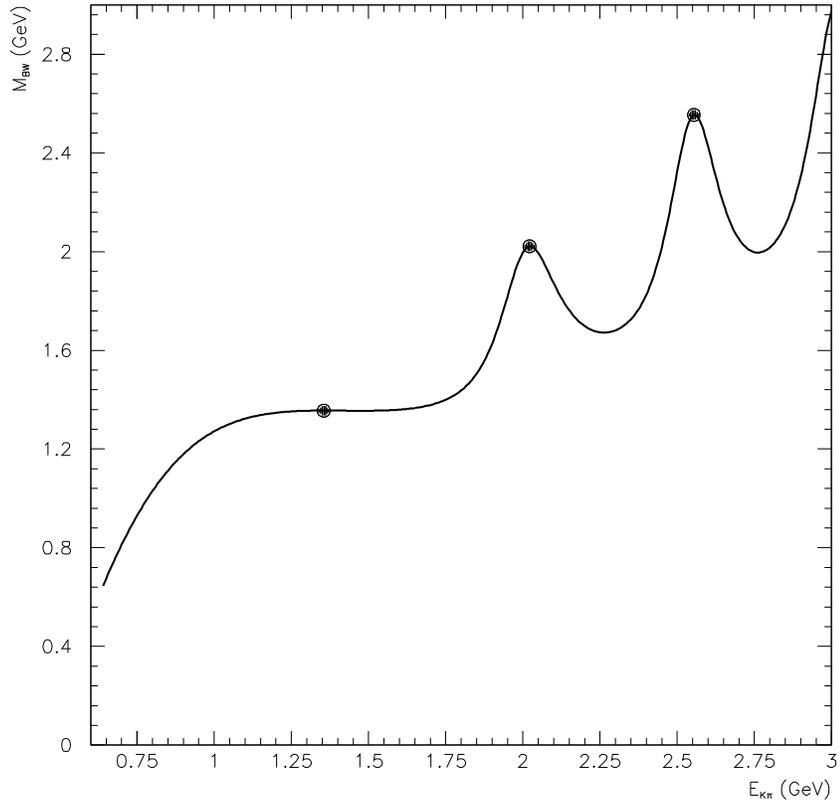} }
\caption
{The dependence of the Breit--Wigner mass as extracted from the
two--channel solution for $K\pi\lra K_0$ scattering.
The dots indicate the resonance masses for the ground, first
radial, and second radial states.
There is a third radial excitation just above 3.0~GeV.
}
\end{figure}

\begin{figure}[htb]
\label{fig:bw_w}
\centerline {\epsfxsize=4.0 true in
\epsffile{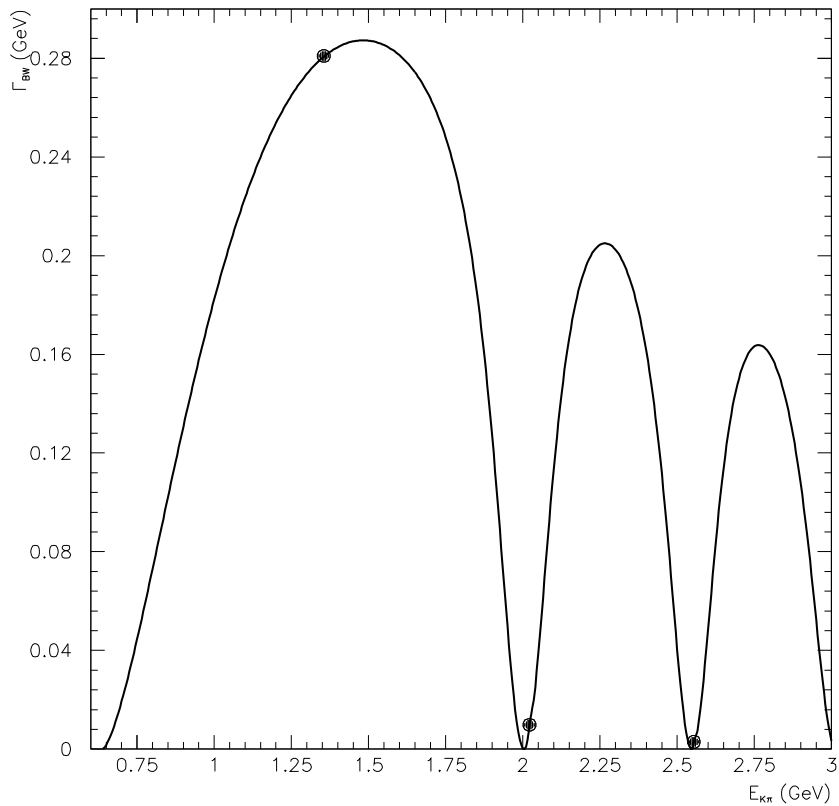} }
\caption
{The dependence of the Breit--Wigner width as extracted from the
two--channel solution for $K\pi\lra K_0$ scattering.
The dots indicate the width predictions at the resonance
energies for the ground, first radially excited, and second
radially excited states.
The location of the dots below the peaks indicates that
the resonances are narrower on the low side than on the high side.
The details of these results will change as more realistic
models for the annihilation potentials are used.
}
\end{figure}

\begin{figure}[htb]
\label{fig:kpi4}
\centerline {\epsfxsize=4.0 true in
\epsffile{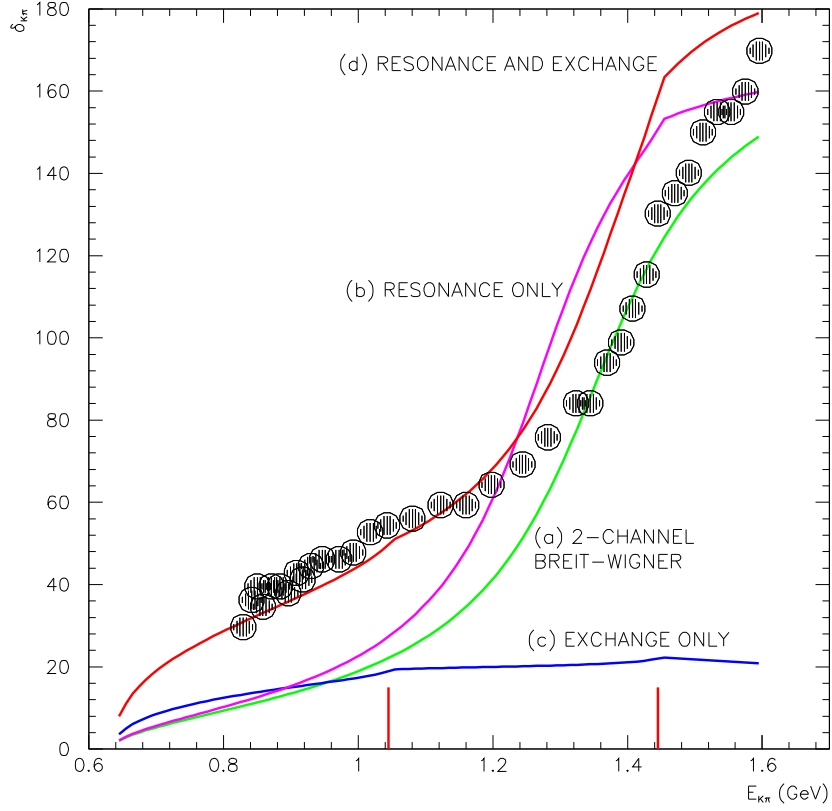} }
\caption {
The $K\pi$ phase shift from the four--channel equation.
(a) uses only $K\pi \lra K_0$ couplings with no quark exchange
potentials (so it's just the Breit--Wigner phase shift),
(b) uses only the resonance couplings of the three two--meson
states to the $K_0$,
(c) uses only quark exchange potentials, and
(d) uses both quark exchange and resonance couplings.
The data is from LASS [20].
The two vertical bars denote the onset of the $K\eta$ threshold
at 1.045~GeV and the $K\eta^\prime$ threshold at 1.445~GeV.
See Section~VI for a discussion of these solutions.
}
\end{figure}

\begin{figure}[htb]
\label{fig:kpim}
\centerline {\epsfxsize=4.0 true in
\epsffile{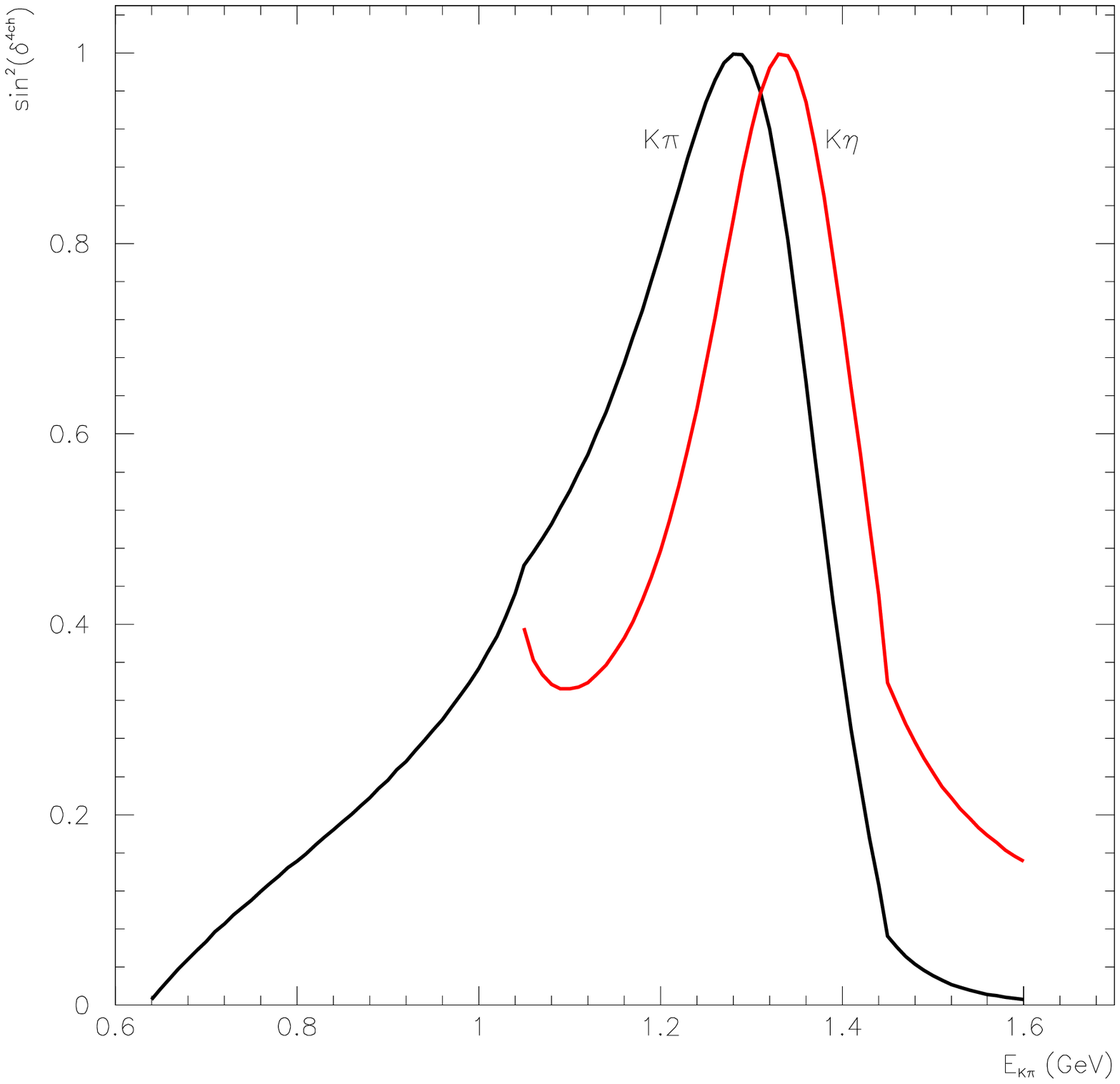} }
\caption {
We plot
$\sin^2(\delta^{4ch}_{K\pi\to K\pi})$ and
$\sin^2(\delta^{4ch}_{K\pi\to K\eta})$
for the scattering processes
$K\pi \to$ $ (K\pi\lra K\eta\lra K\eta^\prime\lra K_0) \to 
K\pi$
and
$K\pi \to$ $ (K\pi\lra K\eta\lra K\eta^\prime\lra K_0)$
$\to K\eta$
where the curves are labelled by their final states.
The 45 MeV mass difference in the central $K_0$ resonance in
these two reactions is a pure multichannel effect.
The decrease in 
$\sin^2(\delta^{4ch}_{K\pi\to K\eta})$ at $K\eta$ threshold
suggests that there is a large $K\eta$ bound state
component in the multichannel system below this energy.
We use both $s$--channel resonance couplings and $t$-channel
quark exchange to generate these curves.
}
\end{figure}

\end{document}